\documentclass[12pt]{iopart}
\usepackage[dvips]{graphicx} 
\usepackage{subfigure}
\usepackage{amssymb}
\usepackage{epsfig}
\usepackage{color}
\usepackage{ifthen}
\usepackage{iopams}  
	
\newcommand{\xx}{{\bf x}}

\def\be{\begin{equation}}
\def\ee{\end{equation}}
\def\bea{\begin{eqnarray}}
\def\eea{\end{eqnarray}}
\def\bn{\begin{enumerate}}
\def\en{\end{enumerate}}
\def\nn{\nonumber}

\def\l{\left}
\def\r{\right}
\def\gnl{g_{\rm NL}}
\def\fnl{f_{\rm NL}}

\begin{document}

\title[Statistical nature of non-Gaussianity from cubic order primordial 
 perturbations\ldots]{Statistical nature of non-Gaussianity from 
cubic order primordial 
 perturbations: CMB map simulations and genus statistic}

\author{Pravabati Chingangbam and Changbom Park}
\ead{prava@kias.re.kr, cbp@kias.re.kr}

\address{Korea Institute for Advanced Study, Hoegiro 87,
  Dongdaemun-gu, Seoul 130722, South Korea. }

\date{\today}

\begin{abstract}
We simulate CMB maps including non-Gaussianity arising from cubic order
perturbations of the primordial gravitational potential, characterized
by the non-linearity parameter $\gnl$. The maps are used to study the 
characteristic nature of the resulting non-Gaussian temperature
fluctuations. We measure the genus and investigate how it deviates
from Gaussian shape as a 
function of $\gnl$ and smoothing scale. We find that the deviation of 
the non-Gaussian genus curve from the Gaussian one has an
antisymmetric, sine function
like shape, implying more hot and more cold  spots for $\gnl>0$ and
less of both for  $\gnl<0$. The deviation increases linearly with
$\gnl$ and also exhibits mild increase as the smoothing scale increases. 
We further study other statistics derived from the
genus, namely, the number of hot spots,
the number of cold spots, combined number of hot and cold spots and the
slope of the genus curve at mean temperature fluctuation. We find
that these observables carry signatures of $\gnl$ that are clearly
distinct  from the quadratic order perturbations, encoded in the
parameter $\fnl$.  Hence they can be very useful tools for distinguishing not
only between non-Gaussian temperature fluctuations and Gaussian ones
but also between $\gnl$ and $\fnl$ type non-Gaussianities.  
\end{abstract}


\maketitle

\section{Introduction}

Since the discovery of the temperature anisotropies of the
cosmic microwave background radiation \cite{cobe:1992} their statistical
nature has been subject of intense study. If inflation 
\cite{guth:1980,Linde:1981mu,Albrecht:1982wi}, as strongly
supported by observations, is indeed the mechanism that gave rise to
these anisotropies, then their statistical nature must be inherited
from those of the primordial density fluctuations. 
All models of inflation, in general,  
predict some amount of deviation of these fluctuations from a Gaussian
distribution. The detailed knowledge of the deviations are quite
model dependent. This makes it a good discriminant between various
models of inflation. The observational search for non-Gaussianity is,
however, beset with serious difficulties since various spurious
observational effects can mask the true signal. 

Primordial non-Gaussianity arises from higher order terms 
in the perturbative expansion of the primordial gravitational
potential, $\Phi$, which must be taken into account in the presence of
higher order interaction. 
In this paper, we consider the following expansion of the {\em local}
type~
\cite{Okamoto:2002ik,Bartolo:2004if,amico:2008}:
\bea
\label{eqn:phi} 
\Phi(\xx) &=& \Phi^L(\xx)+ \fnl \l( (\Phi^L(\xx))^2 - \langle
(\Phi^L)^2 \rangle \r) 
 + \,\gnl (\Phi^L(\xx))^3 + \ldots,
\eea
where $\fnl$ and $\gnl$ are parameters that measure the level of
non-linearity and $\Phi^L$ is the linear order perturbation. 
This expansion is rather special in that it assumes that the higher order 
perturbations are known in 
terms of the linear field, with our ignorance pushed into
$\fnl$ and $\gnl$. Also, they depend only on the linear field value at the
same spatial point. 
Such an expansion is not a generic prediction of
all inflation models. In general, it can be a much more complicated
expression involving convolutions of the products of $\Phi^L$'s ({\em
  non-local}) with the non-linearity parameters being scale dependent
kernels.  
Generally, the predictions of inflation are
quantified by $n$-point functions in Fourier space.
Eq.(\ref{eqn:phi}) holds provided the $n$-point
functions have most of the signal coming from some special 
configurations of the wave vectors.  
For the 3-point function, this corresponds to having the amplitude of 
one of the wave vectors tending to zero,  
the so called {\em squeezed} configuration. For the  4-point function, it 
corresponds to either one of the wave vectors tending to zero 
or two of them tending to 
zero~\cite{Huang:2006eha,Seery:2006vu,Li:2008gg,Chen:2009bc}.  
Here we ignore any possible scale dependence of $\fnl$ and $\gnl$ to
simplify the problem. This is justified by current experimental limitations. 
The bulk of the study on the topic, both theoretical and
observational, has been focused primarily on the quadratic order 
non-Gaussianity~
\cite{Bartolo:2004if,salopek:1990,raghu:1991,Acquaviva:2002ud,maldacena:2002,
Seery:2005wm}. 
The strongest limits on $\fnl$ from CMB observations obtained so far is $-4 <
\fnl < 80$ (at 95\% CL)~\cite{Smith:2009jr}.

There is now growing interest in the cubic order perturbation and
its observability. The question is whether the cubic term is negligible,
comparable or even dominant when compared to the quadratic term. For
the standard single field and uncoupled mutiple fields 
slow-roll inflation they are found to be slow
roll suppressed and hence negligibly small
\cite{Seery:2006vu,Seery:2006js}. (See~\cite{Engel:2008fu} for attempt
to get large trispectrum in single field inflation). 
However, in the curvaton scenario
\cite{linde:1996,lyth:2001} it can happen that the cubic term is
comparable or even dominant to the quadratic term~\cite{sasaki:2006,
 byrnes:2006,kari:2008,Huang:2008bg,Chingangbam:2009xi,Huang:2008zj}. 
Another class of models with similar predictions is the so called multibrid
inflation~\cite{Sasaki:2008uc,Huang:2009xa,Huang:2009vk}. Thus, the 
possibility arises that the
dominant source for primordial non-Gaussianity comes from the cubic
non-linearity and it becomes important to understand theoretically how
observational quantities in the CMB as well as the large scale
structure get affected by it.    
Early studies of the consequences on the CMB have focused on the
angular trispectrum of the temperature fluctuations 
\cite{Okamoto:2002ik,hu:2001,kogo:2006}, while feasibility study of 
measurement of the trispectrum from observational data was performed
in~\cite{troia:2003}. 
Implications of $\gnl$ on galaxy bispectrum was studied
in~\cite{komatsu:bias}. 
The first limit on $\gnl$ using SDSS data
and N-body simulations has been obtained to be $-3.5\times 10^5 < \gnl <
+8.2\times 10^5$  (at 95\% CL) in~\cite{Desjacques:2009jb}. 
No bounds have been obtained using CMB data as yet. 

An important step towards understanding how the
primordial non-Gaussianities affect the  CMB temperature fluctuations
is to simulate maps with the
non-Gaussianities going in as input in the map making process. 
Since we have control over the input parameters, namely, $\fnl$ and
$\gnl$ in this case,  we can test and
calibrate the sensitivities of different statistical observables
to these parameters, using the simulated
maps. Observational contaminants can be added to find out exactly
how each one of them can mask the real non-Gaussian effects and 
experimental bounds can be obtained for the input parameters by comparing
with observational data. 
There are several map making methods that have been proposed in the 
literature for $\fnl$ non-Gaussianity. Komatsu
{\em et al.}~\cite{Komatsu:2003iq} used a straightforward method of
generating a Gaussian random realization of  the linear gravitational
potential in Fourier space and then convolving two such fields to
obtain the quadratic term in Eq.(\ref{eqn:phignl}).  
Liguori {\em et al.}~\cite{liguori:2003} posposed a fast algorithm which
require the computation to be done in real space rather than Fourier space. 
Other methods involve the input of given power spectrum,  
bispectrum and higher order spectra~\cite{Contaldi:2001wr,Rocha:2004ke}, 
or some known correlation
structure of the non-Gaussian field~\cite{vio:2002}, or using sherical
wavelets~\cite{martinez:2002}.  

Simulated non-Gaussian CMB maps arising from $\gnl$ term have not been
discussed in the literature as yet and we present them in this paper.  
The goal is to study how the cubic perturbations 
show up as non-Gaussianity of the temperature anisotropies. We use the
genus, which is the number of isolated hot spots minus the number of
isolated cold spots, as our statistical observable. 
We first make simulations of non-Gaussian maps with $\gnl$
as the input parameter, extending the method of~\cite{liguori:2003}
and using the full linear radiation transfer function.  
In order to be able to investigate the pure effects of the $\gnl$ term
we have set  $\fnl=0$ in our simulations. Then, we use
these maps to compute the genus statistic to find out how it varies as
a function of $\gnl$ and the smoothing scale. 
Further, we discuss four new statistics derived
from the genus and show that they can be very useful tools to
distinguish primordial non-Gaussianity from Gaussianity and also to
distinguish between $\fnl$ and $\gnl$ type non-Gaussianities.  

This paper is organized as follows: in section 2 we outline the method
for generating the non-Gaussian maps, describe the implementation of
the map making process and  we present our
results of the non-Gaussian maps and the one-point PDF. 
In section 3 we compute the genus using the simulated maps
and show how they deviate from the Gaussian shape.  
We then discuss the derived statistics and elaborate on their
characteristics and how they
distinguish $\gnl$ from $\fnl$. We conclude
with a summary of results and remarks on direction for future work 
in Section 4.

\section{Simulation of non-Gaussian maps}

We briefly review the   method for  simulating non-Gaussian maps
outlined in~\cite{liguori:2003} with a simple extension to include
$g_{\rm NL}$ term.

\subsection{Calculating $a_{\ell m}$'s in real space}

The CMB temperature fluctuations are usually expanded in terms of 
spherical harmonics as 
$\Delta T(\hat{n}) = \sum_{lm} a_{lm} Y_{lm}(\hat{n})$. The  $a_{\ell
  m}$'s are then computed by convolving the primordial potential
fluctuations with the radiation transfer function $\Delta_{\ell}(r)$
as,
\begin{equation}
a_{\ell m}  = 4\pi(-i)^l \int \frac{d^3 k} {(2\pi)^3} \,\Phi({\bf k}) \,  
            \Delta_{\ell}(k) \,Y^*_{\ell m}(\hat{\bf k}) \label{alm},
\end{equation}
where $\Phi({\bf k})$ is the Fourier transform of the real space 
potential $\Phi({\bf x})$.  $\Delta_{\ell}(k)$ encodes the
evolution history of the CMB photons in their journey from
recombination till now. 
Defining 
\begin{equation} 
\label{eqn:multipoles_definitions} 
\Phi_{\ell m}(k) \equiv \int \! d\Omega_{\hat{k}} \, 
\Phi(\mathbf{k}) \, Y_{\ell m}(\hat{k}) \; . 
\end{equation}
we can rewrite  $a_{\ell m}$ as 
\begin{equation}
\label{eqn:almk} 
a_{\ell m} =
\frac{(-i)^{\ell}}{2\pi^2} \int \! dk \,\, k^2 \, \Phi_{\ell m}(k) \,
\Delta_{\ell}(k) \; , 
\end{equation} 

To rewrite Eq.~(\ref{eqn:almk}) as an integral in real space, define the real
space harmonic potential 
\begin{equation}
\label{eqn:philmr} 
\Phi_{\ell m}(r) \equiv
\frac{(-i)^{\ell}}{2 \pi^2} \int \! dk \, k^2 \, \Phi_{\ell m}(k) \,
j_\ell(kr) \; , 
\end{equation} 
and its inverse 
\begin{equation}
\label{eqn:philmk} 
\Phi_{\ell m}(k) = 4 \pi (i)^\ell \int
\! dr \, r^2 \, \Phi_{\ell m}(r) \, j_\ell(kr) \; , 
\end{equation} 
where $j_\ell$'s are  spherical Bessel functions. 
Then, insert Eqn.~(\ref{eqn:philmk}) in (\ref{eqn:almk}) 
and define
\begin{equation}
\label{eqn:rtf_definition} 
\Delta_{\ell}(r) \equiv
\frac{2}{\pi} \int \! dk \, k^2 \, \Delta_{\ell}(k) j_\ell(kr) \; ,
\end{equation} 
we can then write:  
\begin{equation}
\label{eqn:almr}
a_{\ell m} = \int \! dr \, r^2 \Phi_{\ell m}(r) \Delta_{\ell}(r) \;. 
\end{equation}

For non-Gaussian $\Phi$ given by Eq.(\ref{eqn:phignl}), we would have 
\begin{equation}
\Phi({\bf k}) = \Phi^{\rm L}({\bf k}) + \fnl \Phi^{\rm NL}({\bf k}) + \gnl
\Phi^{\rm NNL}({\bf k}),
\end{equation}
where 
\begin{eqnarray}
\Phi^{\rm NL}({\bf k}) &=&  \int \frac{d^3k_1}{(2\pi)^3} 
\,\Phi^{\rm L}({\bf k} + {\bf k_1}) \,\Phi^{\rm L}({\bf k_1}), \nonumber \\
\Phi^{\rm NNL}({\bf k}) &=&  \int \frac{d^3k_1}{(2\pi)^3} \frac{d^3k_2}{(2\pi)^3} 
\,\Phi^{\rm L}({\bf k} + {\bf k_1} +  {\bf k_2}) 
 \Phi^{\rm L}({\bf k_1})\,\Phi^{\rm L}({\bf k_2}) . 
\label{eqn:phignl}
\end{eqnarray}
We can then define harmonic components $\Phi_{\ell m}^{\rm NL}$ and
$\Phi_{\ell m}^{\rm NNL}$ in Fourier and real space using
Eqs.~(\ref{eqn:philmr}) and  (\ref{eqn:philmk}), to give us 
\begin{equation}
\Phi_{\ell m}(k)  \equiv \Phi_{\ell m}^{\rm L}(k) + f_{\rm NL} 
\Phi_{\ell m}^{\rm NL}(k) + g_{\rm NL} \Phi_{\ell m}^{\rm NNL}(k). 
\end{equation}
$a_{\ell m}$ is then given by 
\begin{equation}
a_{\ell m} = a_{\ell m}^{\rm L} + f_{\rm NL}\,a_{\ell m}^{\rm NL}+
g_{\rm NL}\,a_{\ell m}^{\rm NNL},
\end{equation}
where each term is an integral over the corresponding $\Phi$. 

Thus, we
need to compute four quantities, namely, $\Delta_{\ell}(r),\ 
\Phi_{\ell m}^{\rm  L}(r),\ \Phi_{\ell m}^{\rm NL}(r) $ 
and $\Phi_{\ell m}^{\rm NNL}(r)
$, in order to get $a_{\ell m}$ upto cubic order primordial perturbations.
$\Delta_{\ell}(r)$ can be independently computed using
$\Delta_{\ell}(k)$ obtained from CMBFAST~\cite{Seljak:1996is}. 
In order to generate $\Phi^L_{\ell m}(r)$ we need its
correlation function given by \cite{liguori:2003},  
\bea
\label{eqn:almrcorr} \left \langle \Phi^{\rm L}_{\ell_1
m_1}(r_1)\Phi^{{\rm L}\star}_{\ell_2 m_2}(r_2) \right \rangle 
&=& \frac{2}{\pi} \delta_{\ell_1}^{\ell_2} \delta_{m_1}^{m_2} \int
\!dk\,k^2 P_\Phi(k) 
 j_{\ell_1}(kr_1) j_{\ell_2}(kr_2) \; , 
\eea 
where $P_\Phi(k)$ is the primordial power spectrum (for the Gaussian
part of $\Phi$), given by
\begin{equation}
P_{\Phi}(k) = \frac{A_0}{k^3}\left(\frac{k}{k_0}\right)^{n_s-1},
\end{equation} 
with $A_0$ being the amplitude, $k_0$ is some suitable pivot scale and
$n_s$ is the spectral index.  Then,
$\Phi^L_{\ell m}(r)$ can be obtained from the integral
\begin{equation}
\label{eqn:real_convolution} 
\Phi^{\rm L}_{\ell m}(r) = \int \! dr_1 \, r_1^2 \, n_{\ell m}(r_1) 
W_\ell(r,r_1) \; ,
\end{equation} 
where $n_{\ell m}(r)$ are independent complex Gaussian variables 
characterized by the correlation function 
\begin{equation}
\label{eqn:whitenoise} 
\left \langle n_{\ell_1 m_1}(r_1)
n^*_{\ell_2 m_2}(r_2) \right \rangle = 
\frac{\delta^D(r_1-r_2)}{r^2}\delta_{\ell_1}^{\ell_2} \delta_{m_1}^{m_2}\; ;  
\end{equation} 
and $W_\ell(r,r_1)$ are filter functions defined as
\begin{equation}
\label{eqn:filter} 
W_\ell(r,r_1) =
\frac{2}{\pi} \int \! dk \, k^2 \, \sqrt{P_\Phi(k)} \, j_\ell(kr)
j_\ell(kr_1) \; .  
\end{equation} 
For fixed $r$,  $W_\ell(r,r_1)$ is a smooth function of $r_1$ and 
sharply peaked at $r=r_1$. Simplified expressions which are convenient
for numerical calculation of $W_\ell(r,r_1)$ are given in 
Appendix A. 

To compute $\Phi^{\rm NL}_{\ell m}$, first  compute the linear potential 
$\Phi^{\rm L}({\bf r}) = \sum_{\ell m} \Phi^{\rm L}_{\ell m}(r) Y_{\ell m}
({\hat r})$ and square it to obtain  $\Phi^{\rm
  NL}({\bf r})$. Then harmonic transform to get  
$\Phi^{\rm NL}_{\ell m}(r)$. 
Similarly, $\Phi^{\rm NNL}_{\ell m}$, can be computed by first taking
  cube of $\Phi^{\rm L}({\bf r})$  to obtain 
$\Phi^{\rm NNL}({\bf r})$ and then harmonic transforming to
  get 
$\Phi^{\rm NNL}_{\ell m}(r)$. Finally, putting $\fnl=0$ we get,
\begin{equation}
\Delta T = \Delta T^{\rm G} + \gnl \Delta T^{\rm NG}. 
\end{equation}
This method is particularly useful for
  calculating the $\gnl$ case because had we done
  the calculation in $k$ space we would have had to convolve three
  $\Phi^L$'s as in Eqn.(\ref{eqn:phignl}).

\begin{figure}
\begin{center}
\includegraphics[height=0.3\textheight,width=.45\textwidth]{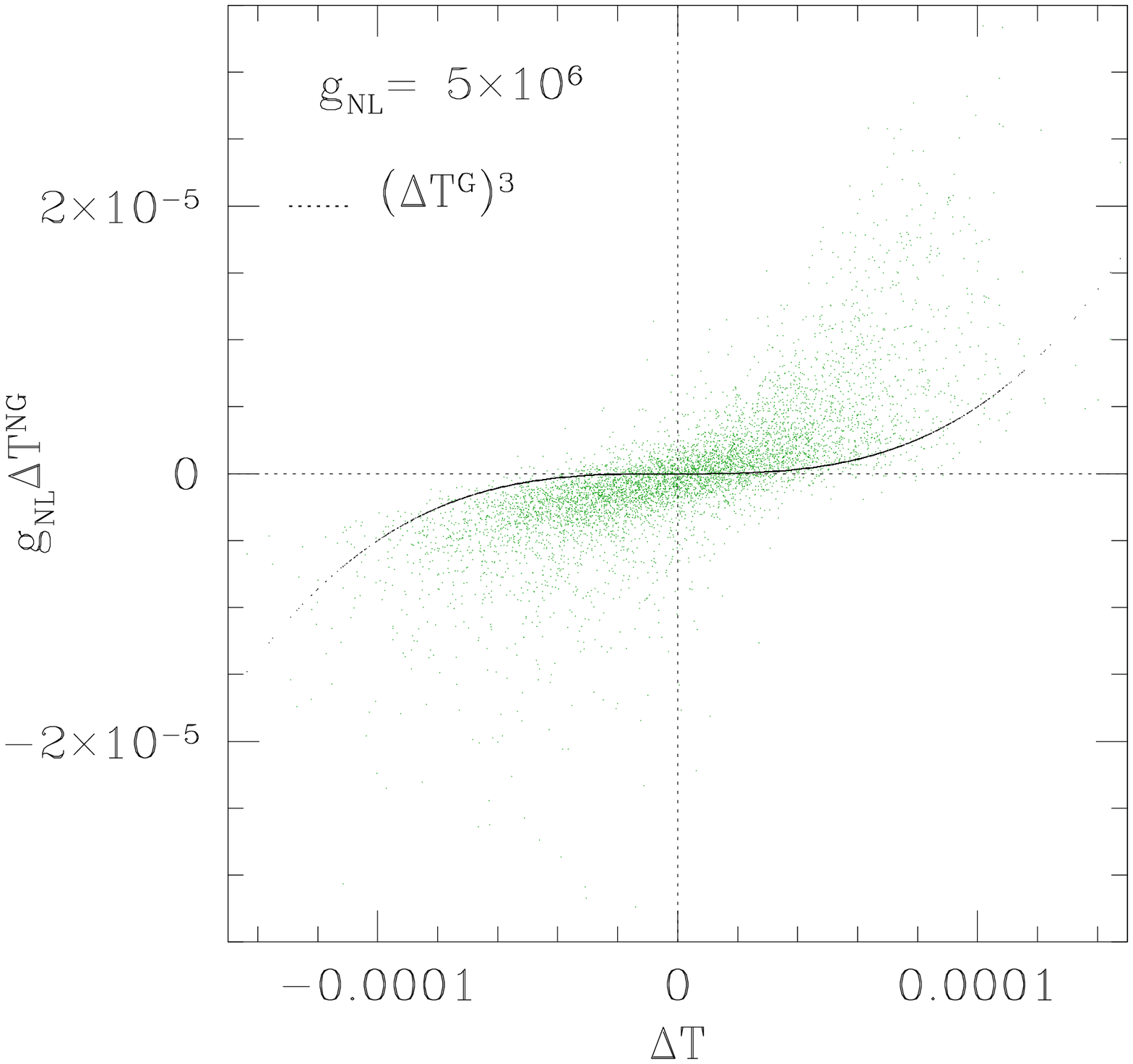}\quad 
\includegraphics[height=0.3\textheight,width=.45\textwidth]{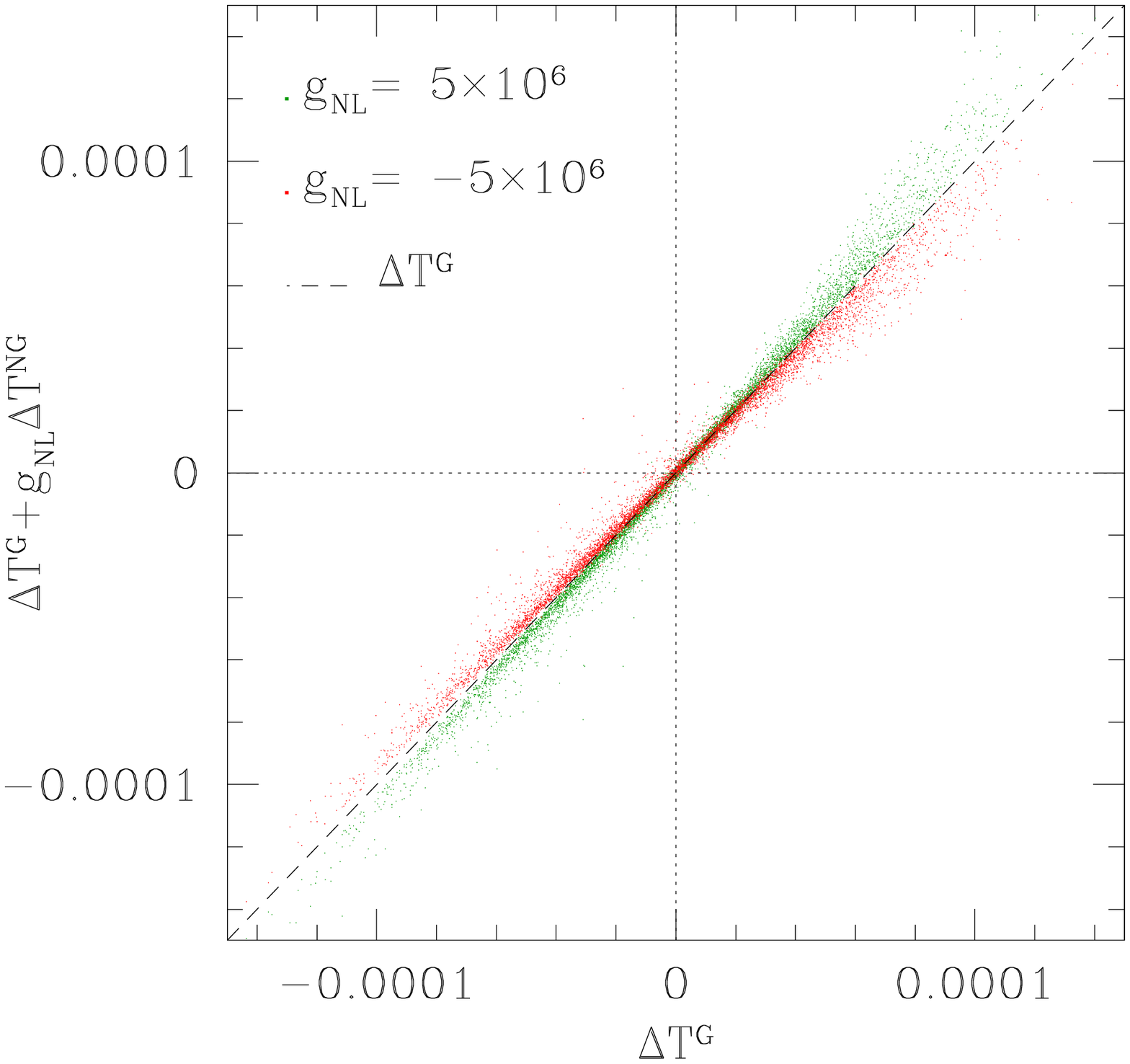}
\caption{Pixel Distribution of $\Delta T^{\rm NG}$ with respect to 
 $\Delta T^{\rm G}$.}
\label{fig:dt}    
\end{center}
\end{figure} 

\subsection{Implementation of the algorithm}

\begin{figure*}
\centering
\subfigure{\includegraphics[height=0.21\textheight,width=0.46\textwidth]
{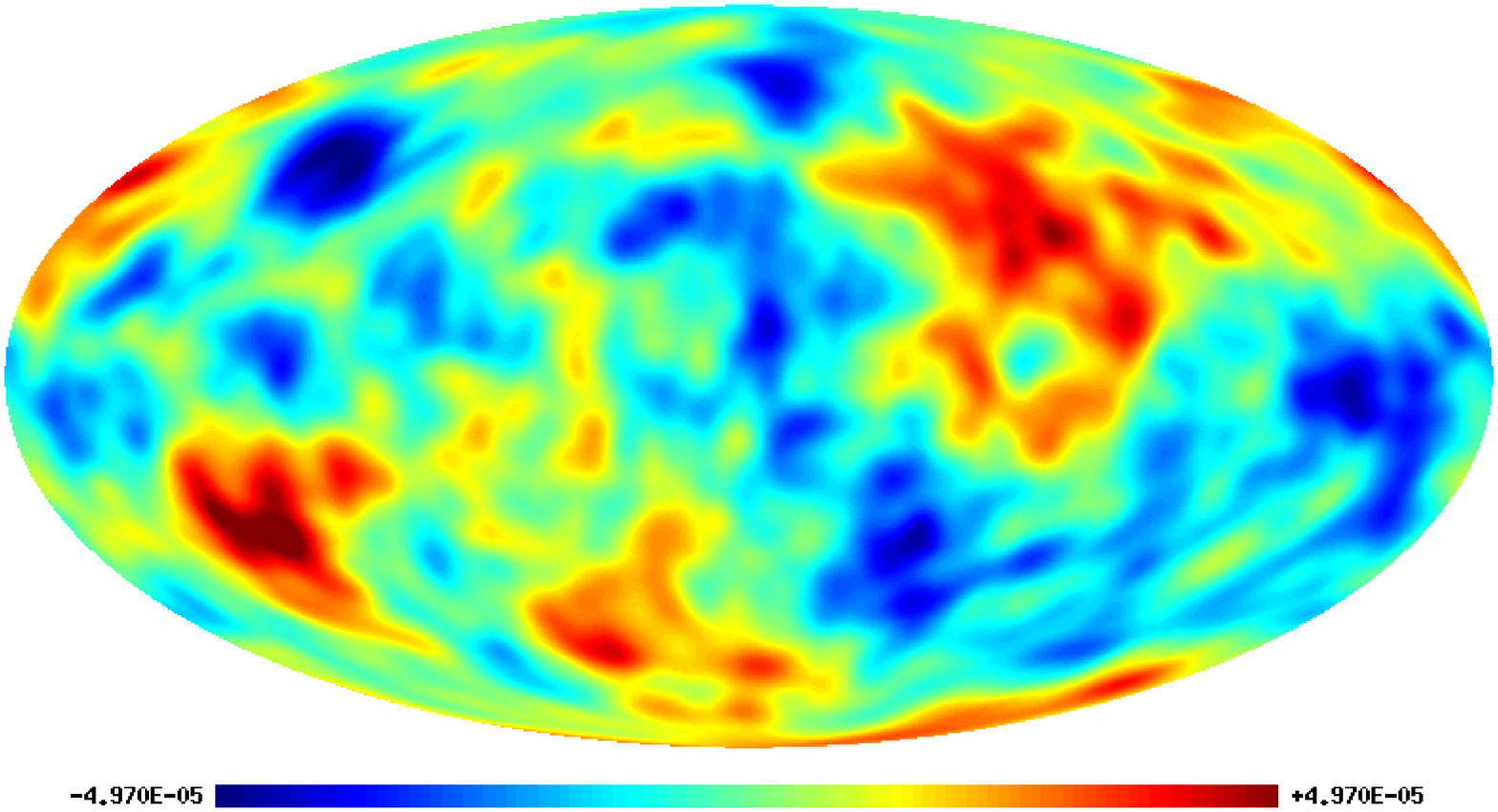}} \qquad
\subfigure{\includegraphics[height=0.21\textheight,width=0.46\textwidth]
{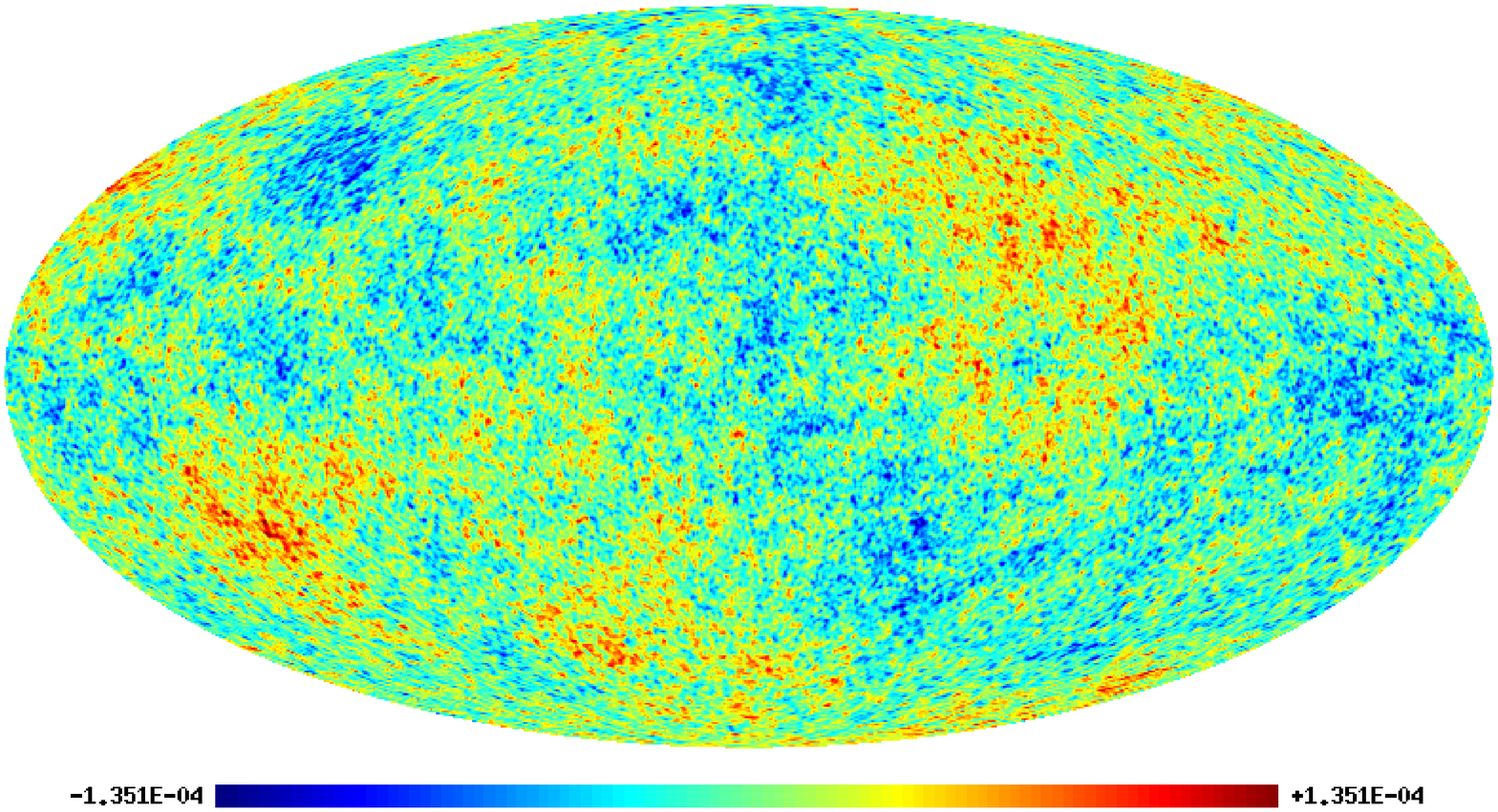}} \\
\subfigure{\includegraphics[height=0.21\textheight,width=0.46\textwidth]
{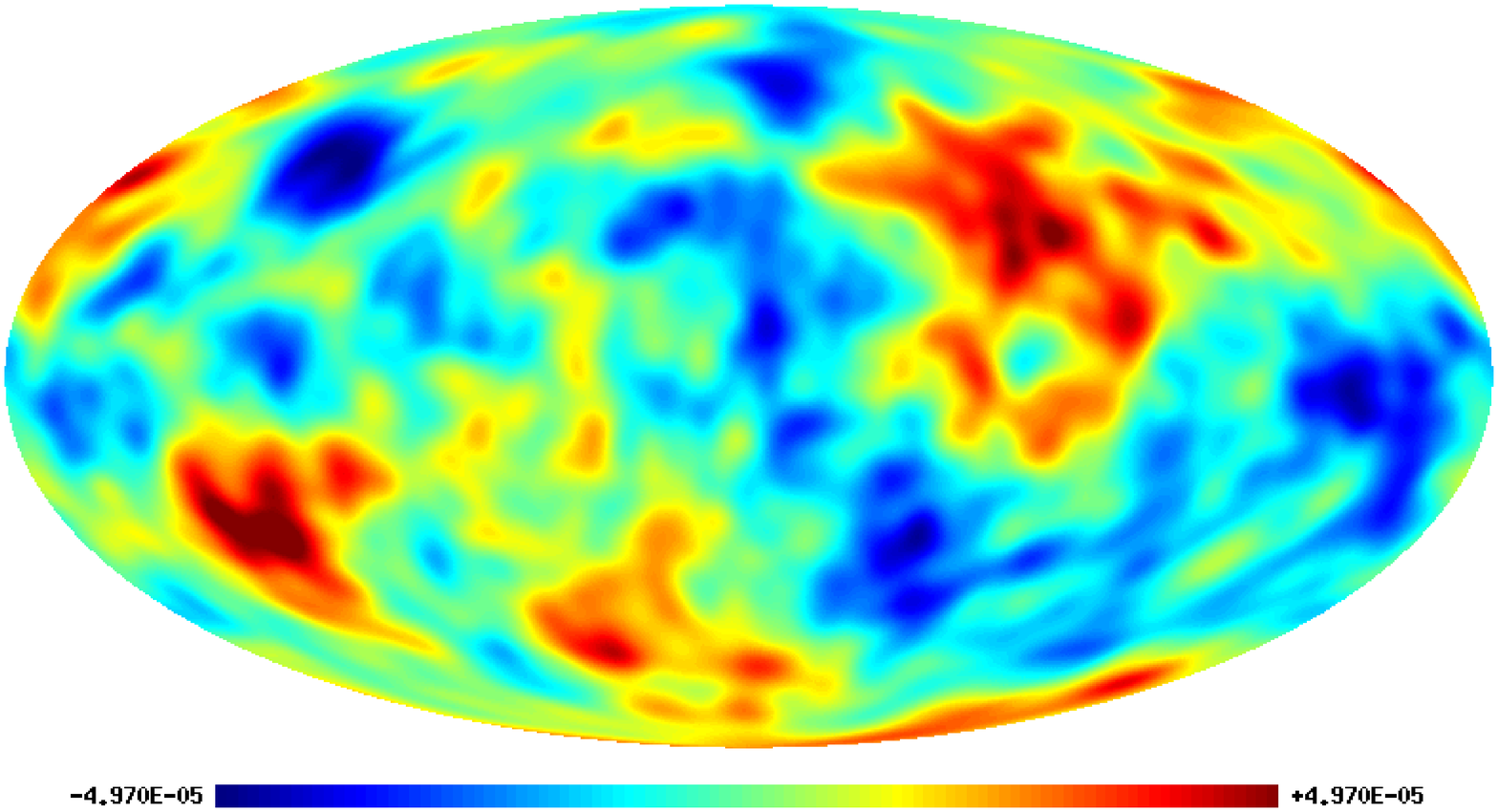}} \qquad
\subfigure{\includegraphics[height=0.21\textheight,width=0.46\textwidth]
{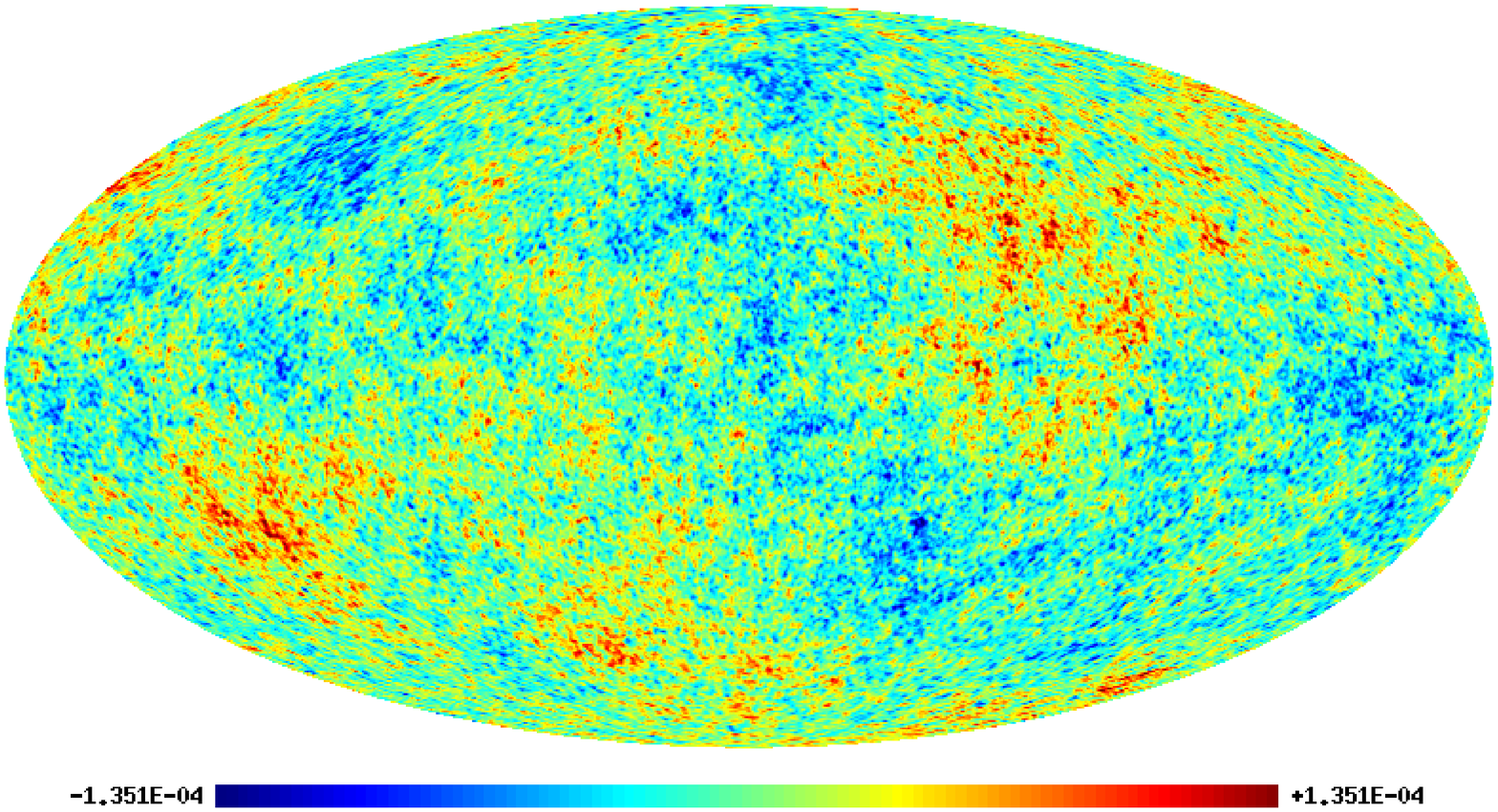}} \\
\subfigure{\includegraphics[height=0.21\textheight,width=0.46\textwidth]
{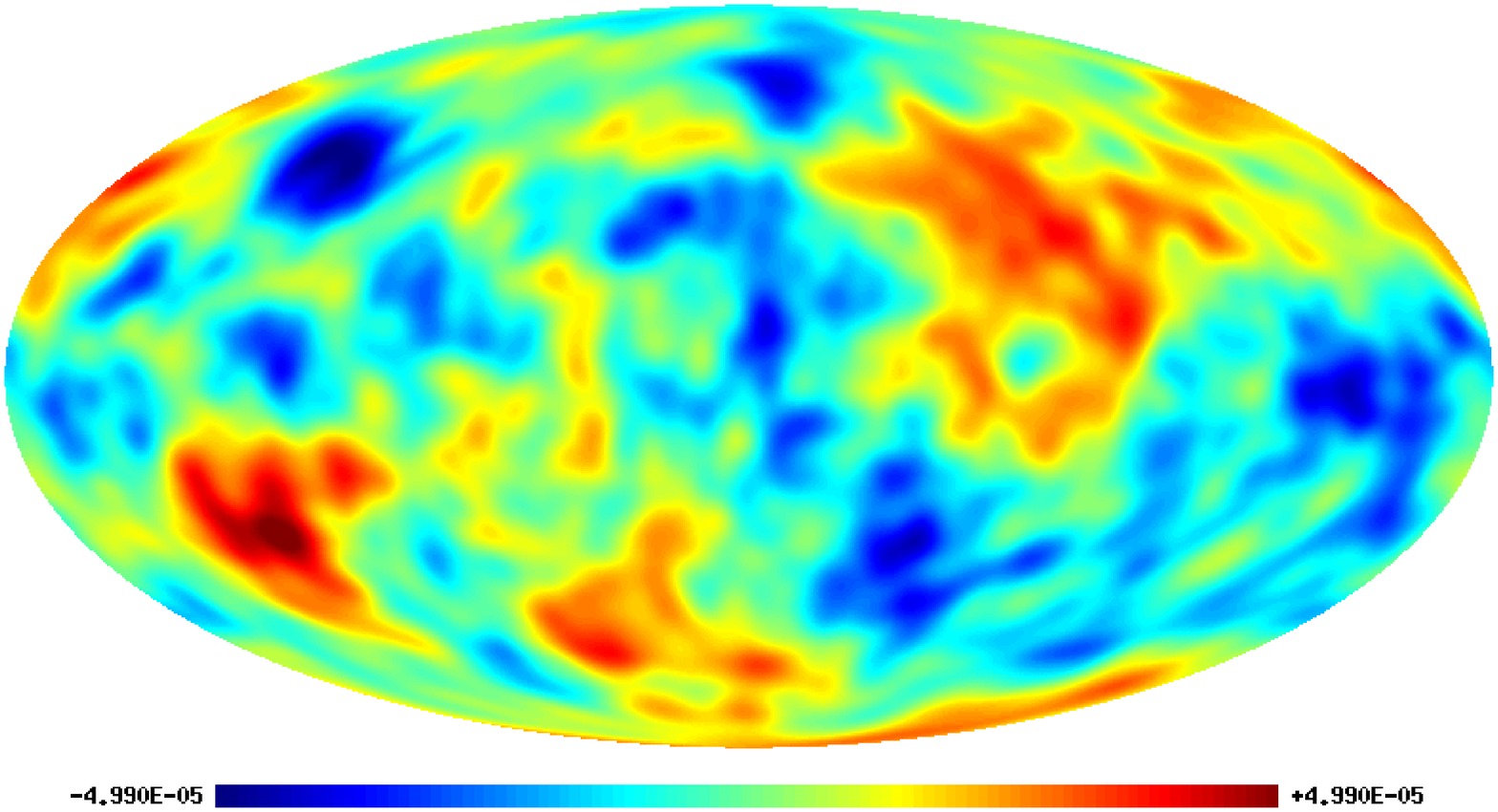}} \qquad
\subfigure{\includegraphics[height=0.21\textheight,width=0.46\textwidth]
{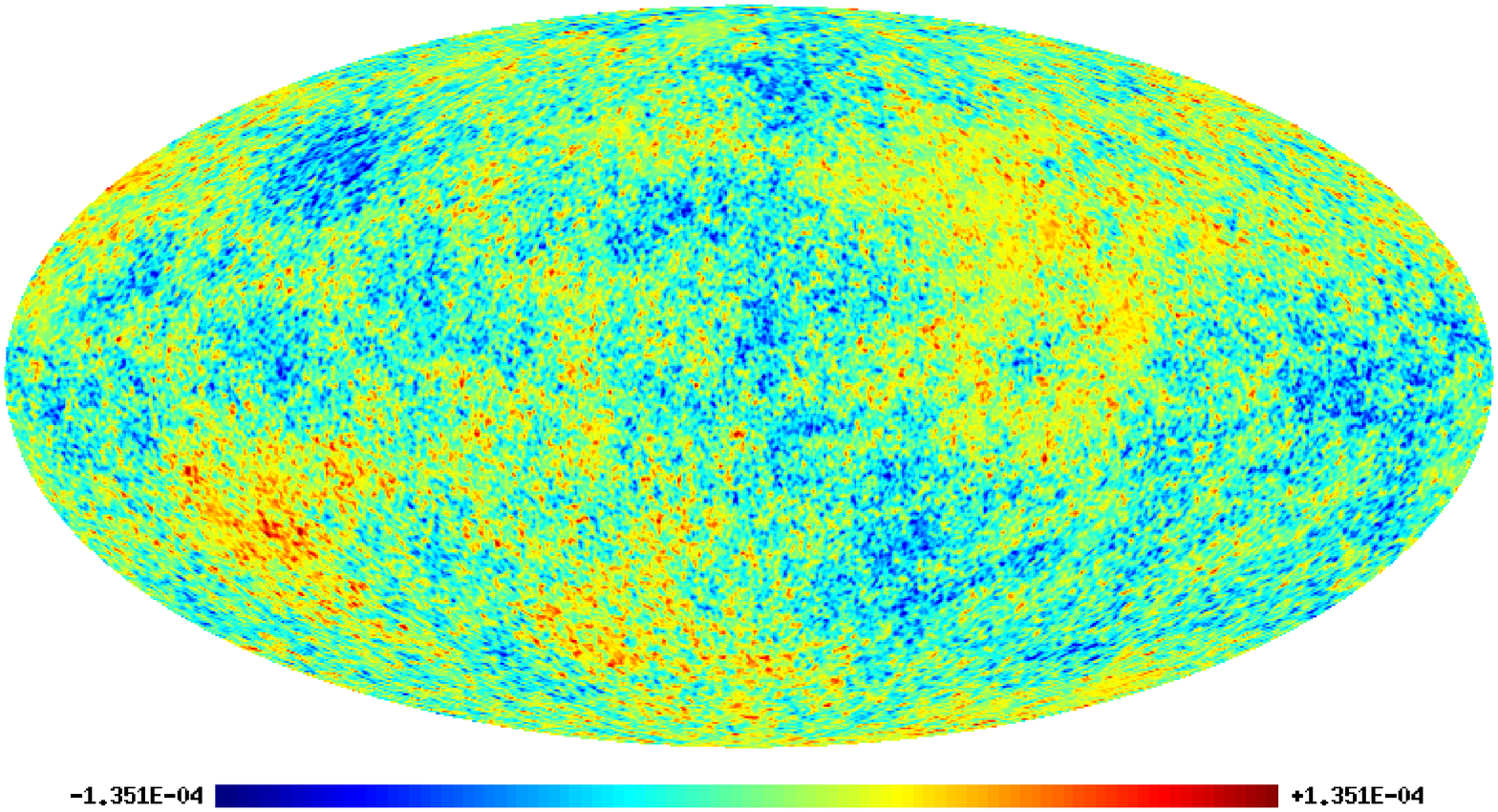}}
\caption{All maps in this figure are obtained from one Gaussian
  realization. The left-hand ones are smoothed  by   
${\rm FWHM}=7^\circ$ while
  the right hand ones are smoothed by ${\rm FWHM}=30'$. The top
  ones are Gaussian, middle are non-Gaussian with $\gnl=5\times 10^{6}$ 
and the bottom ones have $\gnl=- 5\times 10^{6}$.}  
\label{fig:maps}    
\end{figure*}

We use a $\Lambda$CDM cosmological model with primordial
spectral index $n_s=1$.  
We have used the WMAP 5 year parameters~\cite{komatsu:2008} given by  
$\Omega_c=0.233,\ \Omega_b=0.0462,\  
\Omega_{\Lambda}= 0.721,\ \tau_{re}= 0.087,\ h_0=0.719 $. 
With these parameters, the conformal time today is $\tau_0=14360
\,{\rm Mpc}^{-1}$. 
The accuracy of $\Phi^L_{\ell m}(r)$ can be
tested by computing its `angular power spectrum' \cite{liguori:2007}, 
which for $n_s=1$ is obtained as 
$\frac{1}{2\ell+1} \,\sum_m \,|\Phi^{\rm L}_{\ell m}(r)|^2
\propto 1/\ell(\ell+1) $. 
We have sampled $r$ at $472$ points, with different
step sizes chosen  at different epochs, based on the shape of
$\Delta_{\ell}(r)$.  
The accuracy of the resulting Gaussian $C_{\ell}$
was tested by averaging over several maps and then comparing with the
theoretical output of CMBFAST. 
We have used the
Healpix package~\cite{Gorski:2004by} to perform the harmonic transforms 
and the CMBFAST package~\cite{Seljak:1996is} to compute the radiation
transfer  functions $\Delta_{\ell}(k)$. 
We have fixed $\ell_{\rm max}=1100$ and the harmonic 
transforms have been computed using $N_{\rm side} = 512$. 
We find  that the $\gnl$ term begins to dominate the linear term
roughly around $\gnl\sim 10^7$ and hence the perturbation expansion of 
$\Phi$ is invalid beyond this value.    
The Gaussian maps are normalized by CMBFAST, while  
the non-Gaussian ones are normalized by matching the values of
$C_{\ell}$ at the first acoustic peak, $\ell=220$, with the Gaussian one. 

In Fig.(\ref{fig:dt}) we show for one simulation, how for a
given pixel  $\gnl \Delta T^{\rm NG}$ and  $\Delta T^{\rm G}$ are
correlated. Each of the green and blue dots represents a
pixel. The left panel shows the pixel distribution of pure  non-Gaussian 
temperature $\Delta T^{\rm NG}$ about the $(\Delta
T^{\rm G})^3$ curve. The right panel shows how the full $\Delta T $,
deviates from the $\Delta T^{\rm G}$ line, for positive and negative
values of $\gnl$.

\subsection{Non-Gaussian maps and 1-point PDF}

\begin{figure*}
\centering
\subfigure{\includegraphics[height=0.3\textheight,width=0.45\textwidth]
{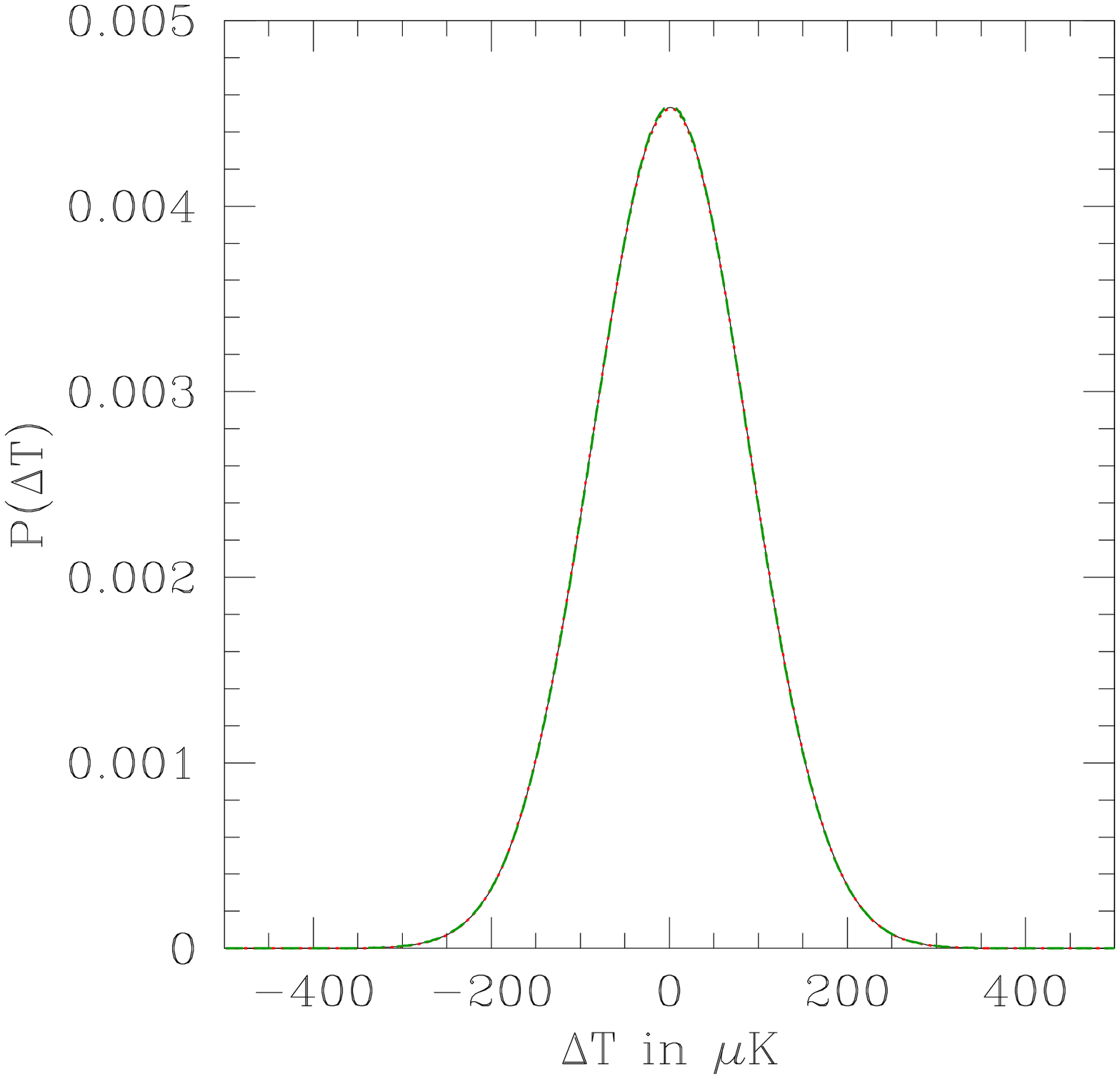}} \hskip .2cm
\subfigure{\includegraphics[height=0.3\textheight,width=0.45\textwidth]
{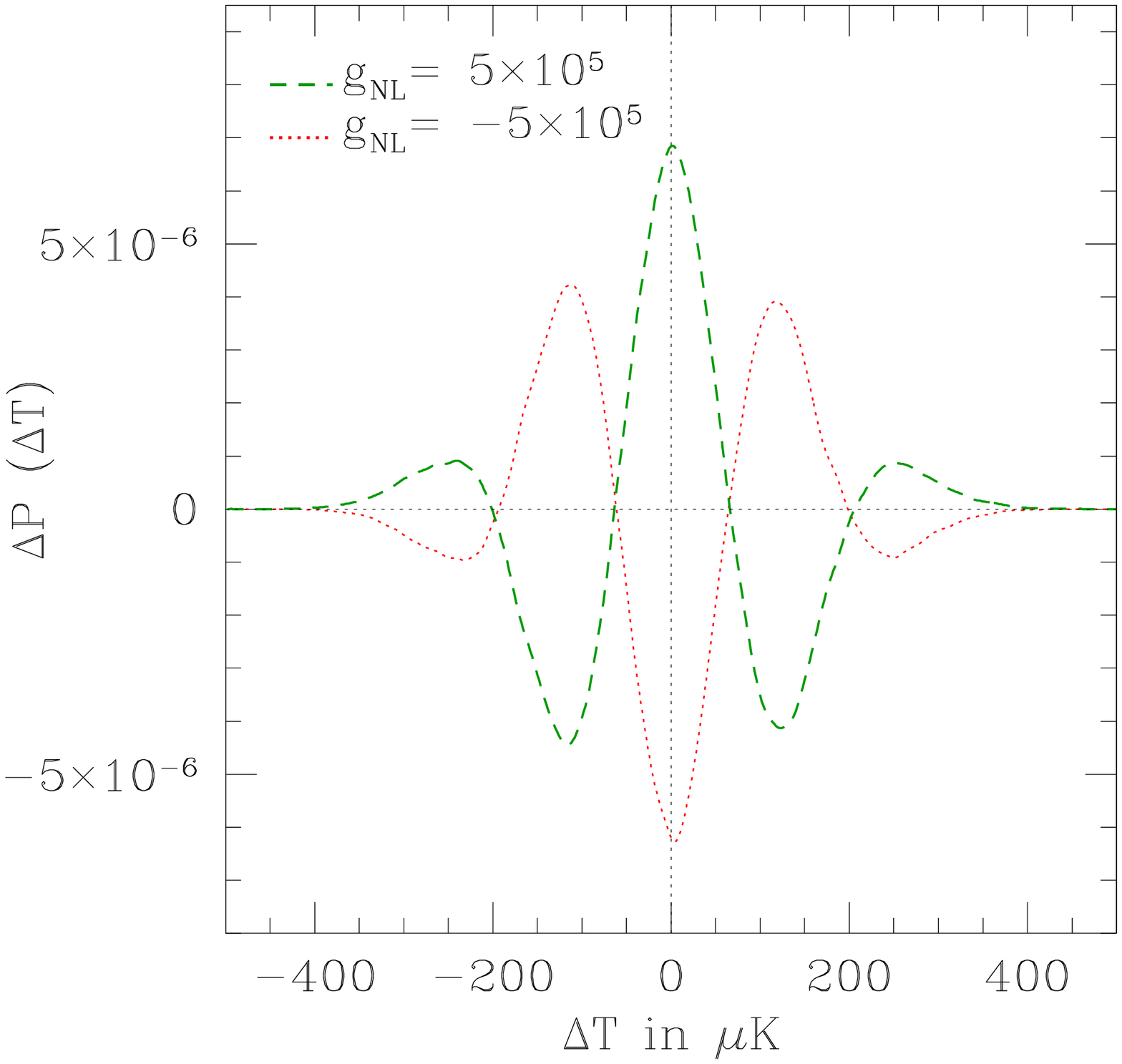}} 
\caption{Effect of $\gnl$ on 1-point PDF. Black (solid) line correspond to 
Gaussian, while green (dashed) corresponds to non-Gaussian with
$\gnl >0$ and red (dotted)  corresponds to $\gnl<0$. Results are averaged over
  150 Gaussian and non-Gaussian maps smoothed with FWHM=30'. 
The right panel shows the deviation of $P(\Delta T)$ of the
non-Gaussian maps from the Gaussian ones.} 
\label{fig:histo}    
\end{figure*} 

Gaussian and non-Gaussian maps obtained for the same Gaussian
realization, for two different Gaussian 
smoothing scales, are shown in~Fig.~(\ref{fig:maps}). 
The positive $\gnl$ maps show hot spots
that are relatively hotter than those in the Gaussian map. On the
other hand, the maps with negative $\gnl$ show relatively cooler
hotspots. The large $\gnl$ value $\pm 5\times 10^6$ 
is chosen to make the differences in
the maps visible. Note that the maps vary from realization to realization due
to statistical fluctuations and at such large value of $\gnl$ the
non-Gaussian term of the temperature fluctuation may dominate over the
Gaussian part for some realizations. We have shown here a realization 
which is close to the average behavior indicated by the average 1-point PDF
described below and for which the non-Gaussian term is still
sub-dominant to the Gaussian term despite the large value of $\gnl$.    

The 1-point PDF's,  $P(\Delta T)$,  are plotted in~Fig.~(\ref{fig:histo}) for 
$\gnl=\pm 5\times 10^5$ and FWHM=30', averaged over 150 realizations. 
$\Delta P$ is the difference between  non-Gaussian and  Gaussian PDF's
and we have shown them in the right hand side
of~Fig.~(\ref{fig:histo}) for positive and negative $\gnl$'s. We see that 
for both positive and negative $\gnl$, the mean position is not
affected, as expected from the fact that non-Gaussian part comes from
$\left( \Phi^{\rm L}\right) ^3$. Positive $\gnl$ increases the pixels
around the mean temperature fluctuations, decreases the intermediate
temperature range and again increases the hottest and coldest ranges,  
leading to leptokurtic shape of the 1-point PDF relative to the
Gaussian one.  Negative $\gnl$ has the opposite effect and results in
platykurtic shape of the 1-point PDF relative to the Gaussian
one. These effects become more pronounced as we increase $\gnl$ and
corroborates what we observe visually in the maps shown
in~Fig.~(\ref{fig:maps}). On the angular power spectrum, $C_{\ell}$, 
both positive and negative $\gnl$ increase the low scale power with
negative $\gnl$ having a stronger effect.


\section{Genus statistic}

\begin{figure*}
\centering
\subfigure{\includegraphics[height=0.56\textheight,width=1.\textwidth]
{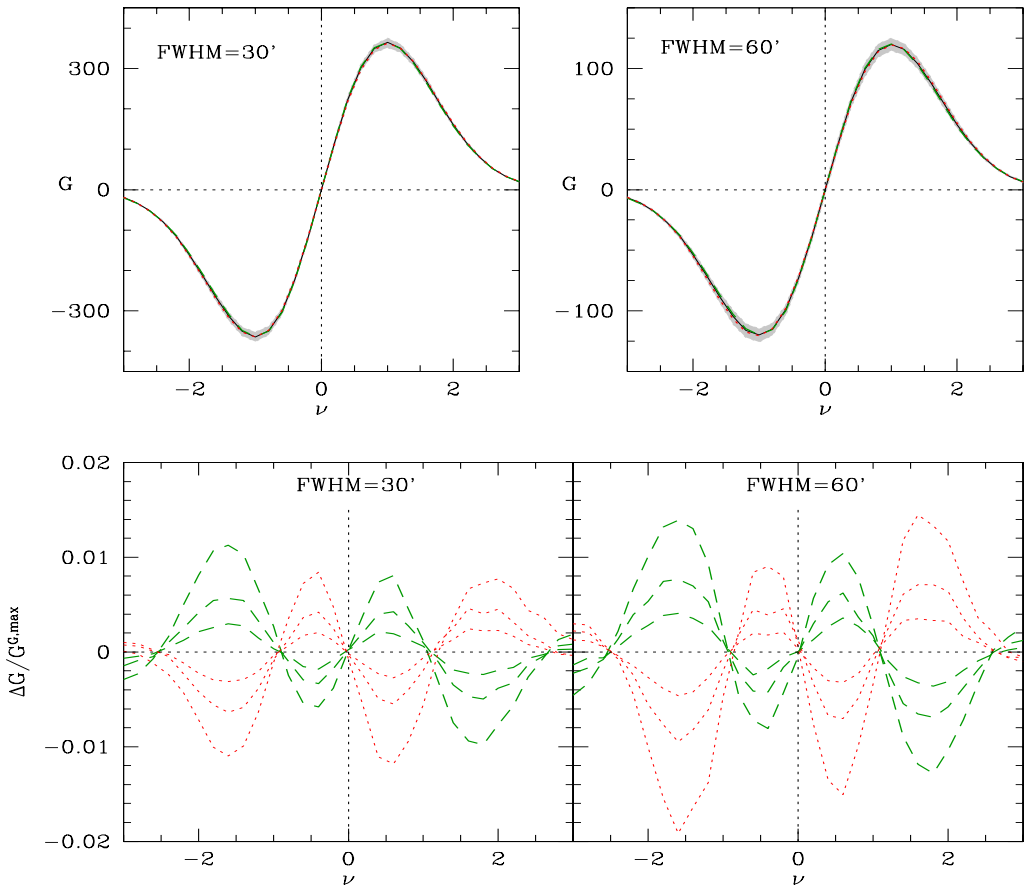}}
\caption{Variation of $g(\nu)$ with $\gnl$ for different
  smoothing angles . Results have been averaged
  over 200 realizations. $G^{G,{\rm max}}$ is the amplitude of the
  Gaussian genus evaluated at $\nu=1$. 
  Black (solid) lines are Gaussian, green (dashed) lines
  correspond to postive $\gnl$ and red (dotted) to negative $\gnl$. 
The values of $\gnl$ are $\pm 5\times 10^5,\, \pm 1\times 10^{6}$ and
  $\pm 2\times 10^{6}$. The shaded regions in the top panels show the
  1-$\sigma$ error bars for the Gaussian genus curve.}
\label{fig:pnge}    
\end{figure*}

By means of iso-temperature contours of the temperature fluctuation
field one can study its global morphological properties. The genus,
which is the number of isolated hot spots minus the number of isolated
cold spots, can be obtained from the contours for a given
threshold temperature, denoted by $\nu\equiv \Delta T/\sigma_0$, where
$\sigma_0$ is the standard deviation of the temperature fluctuation. 
It is sensitive to the Gaussian/non-Gaussian
nature of the fluctuation field and the shape of the underlying
angular power spectrum. This makes it a useful tool to test
non-Gaussianity.  It was introduced in
the context of the CMB in the seminal
papers~\cite{coles:1988,gott:1990}. It has been used extensively to
study non-Gaussianity in a number of 
papers~\cite{changbom:1997,Colley:2003sp,changyung:2004}. 
The genus is one of the three Minkowski
functionals (MF's)~\cite{gott:1990,gorski:1998,Winitzki:1997jj}, which are
topological quantities that can be defined for a two dimensional field
and which completely characterize its topological properties. The
other two are the total iso-temperature contour length and the
fraction of total area above the threshold. They have been 
used to constrain the $\fnl$
parameter~\cite{komatsu:2008,Gott:2006za,Hikage:2008gy,Natoli:2009wk}.  

For a given temperature threshold the genus is given by
\begin{equation}
G(\nu) = \frac{1}{2\pi} \int_C \kappa ds, 
\end{equation}
where $\kappa$ is the signed curvature of the iso-temperature contours
$C$. The genus can also be parametrized by the area fraction above the
threshold. Using the temperature threshold is computationally easier, 
while the area fraction decreases correlations between the
MF's~\cite{shandarin:2002}.  Here we use the 
temperature threshold since we are focussing on the genus only. 

For a given $C_{\ell}$ and Gaussian smoothing angle $\theta_s$, related to
${\rm FWHM}$ as $\theta_s={\rm FWHM}/2\sqrt{2\ln 2}$, the genus per
steradian of a Gaussian temperature field can be expressed as
\begin{equation}
G(\nu) = \frac{1}{2(2\pi)^{3/2}} \frac{\sum \ell(\ell+1)(2\ell+1)
  C_{\ell}  F_{\ell}^2}{\sum (2\ell+1) C_{\ell} F_{\ell}^2} \
 \nu e^{-\nu^2/2}, 
\end{equation}
where $F_{\ell}$ is the smoothing filter, which for a Gaussian filter
is given by $F_{\ell}= \exp(-\ell(\ell+1)\theta_s^2/2)$.  
For a weakly
non-Gaussian field characterized by $\fnl$, approximate analytic
expressions for the Minkowski functionals were obtained
in~\cite{Matsubara:2003yt,Hikage:2006fe}. The expressions upto $\gnl$ order
on the perturbations have not been calculated yet. 
Here we study the topology of the temperature
fluctuations that arise purely from $\gnl$ non-Gaussianity 
by measuring the genus.
For the computation of the genus we follow the method
of~\cite{gott:1990}. We have used 31 threshold levels in the range -3
to 3. The amplitude and shape of the genus curve is sensitive to the 
smoothing scale and we 
have chosen three scales, ${\rm FWHM}=30',\,45'$ and $60'$ to
demonstrate our results. Results become more noisy at higher smoothing
scales since the number of structures (hot and cold spots) decrease.    
We have averaged the results over 200 maps for each
value of $\gnl$.

\begin{figure}
\begin{center}
\includegraphics[height=0.3\textheight,width=.45\textwidth]{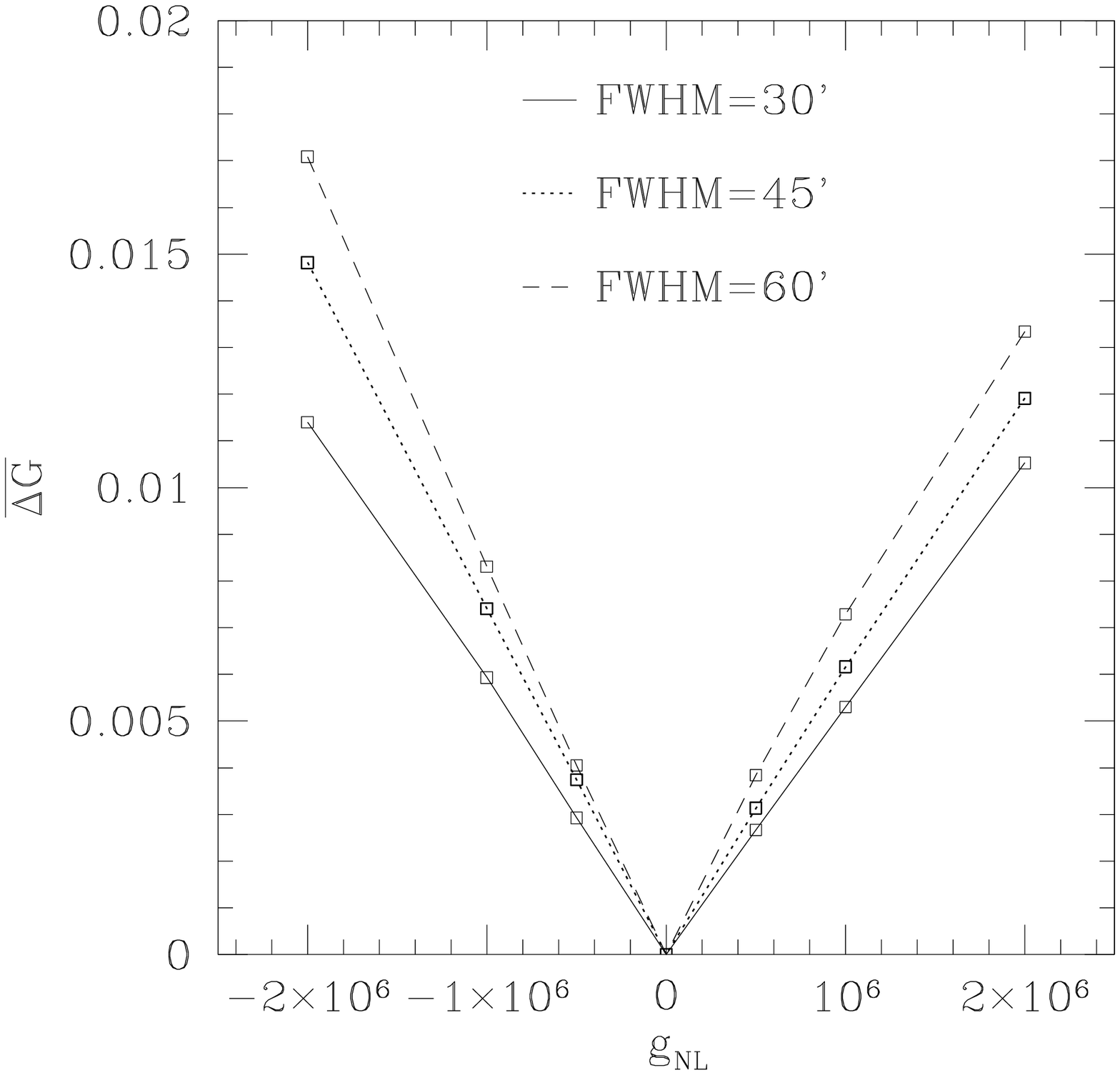}
\quad
\includegraphics[height=0.3\textheight,width=.45\textwidth]{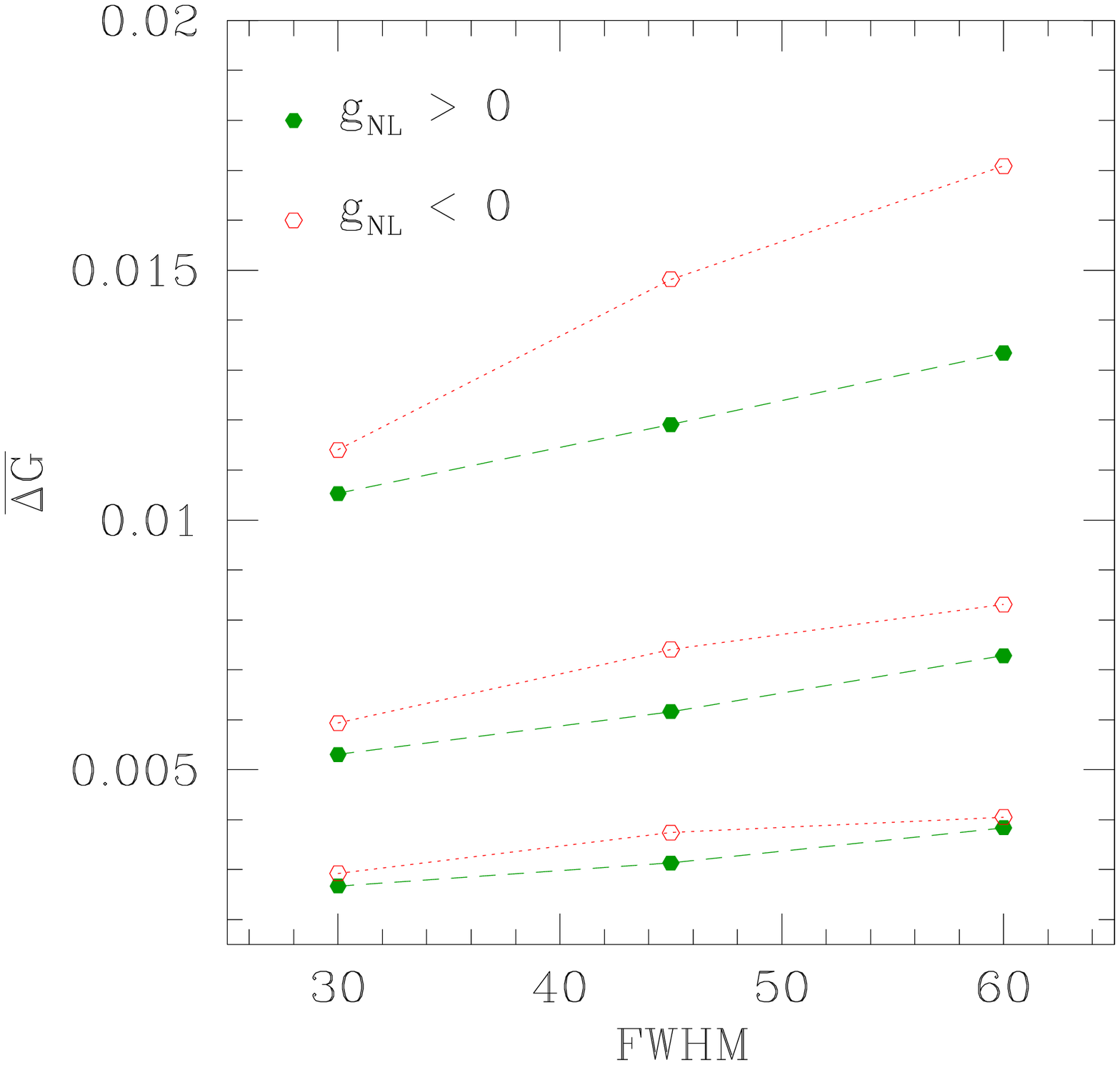}
\caption{Functional dependence of $\Delta G/G^{G,{\rm max}}$ on $\gnl$
  and smoothing scale. In the right hand figure, the lines from the bottom
  to the top correspond to $\gnl=\pm5\times 10^5, \pm10^6,\pm2\times 10^6 $ .}
 \label{fig:dgbygfunc}    
\end{center}
\end{figure}

Let us denote: 
\begin{equation}
\Delta G(\nu) \equiv G^{NG}(\nu) -  G^G(\nu),
\end{equation}
where $G^G(\nu)$ is the Gaussian genus and  $G^{NG}(\nu)$ is the
non-Gaussian one. In Fig.~(\ref{fig:pnge}) we have plotted $G$ and 
$\Delta G/G^{G,{\rm max}}$, where $G^{G,{\rm max}}$ is the amplitude
of the Gaussian genus at $\nu=1$. We have shaded the region within
1-$\sigma$ error bars for the Gaussian genus curves in the upper panels. 
It shows that for positive $\gnl$, the amplitude of the non-Gaussian
 genus curve is higher than the Gaussian one in the threshold range 
 $0\lesssim |\nu| \lesssim 1$, while it is lower in the
 range $1\lesssim |\nu| \lesssim  2.5$. The fact that the genus is
 smaller in  the range $1\lesssim |\nu| \lesssim  2.5$ means  there
 are fewer hot 
 spots and cold spots in the CMB map when non-Gaussian contribution
 with positive $\gnl$ is present.  
Because of the larger genus amplitude at  $0\lesssim |\nu| \lesssim
1$, the range of $\nu$ showing the sponge-like
topology~\cite{Gott:1986uz} (actually a 
two dimensional cut through sponge in the context of CMB) is smaller
relative to the Gaussian one.  When $\gnl$ is negative, there are
 more hot as well as cold spots. Other than an overall scaling of the 
 amplitude of $\Delta G$, the smoothing scale does not seem to affect 
its shape in the threshold range $1\lesssim |\nu| \lesssim  2.5$.

In order to quantify the functional dependence of $\Delta G/G^{G,{\rm 
max}}$ we take the average of the magnitudes of its peak values 
(which lie roughly in the ranges $1.5< \nu <2$ and $-2 <\nu<-1.5$). 
Let us denote it by
$\overline{\Delta G}$. In the left panel of Fig.~(\ref{fig:dgbygfunc})
we have plotted the variation of $\overline{\Delta G}$ with $\gnl$ for
FWHM  $30', 45'$ and $60'$. We find a linear dependence on $\gnl$, 
indicating that the 
leading order contribution to the deviation from the Gaussian genus
curve comes from terms of order $\gnl$. The right panel 
shows the  variation of $\overline{\Delta G}$ with the
smoothing scale for fixed $\gnl$. We find a mild increase as we
increase FWHM, in the range of smoothing scales that we have studied.  

So far we have described the number of structures for threshold range 
$|\nu| < 2.5$.  Upto the smoothing smoothing scale of FWHM$=60'$ the
number of structures above this threshold are not significantly large.  
However, an interesting observation that we have made for smoothing
scales above FWHM$=60'$ is that the relative number of structures in
this hottest or coldest range of threshold values, $|\nu| > 2.5$, 
compared to the range $|\nu| < 2.5$, grows significantly and hence
promises to be useful for constraining $\gnl$ at higher smoothing
scales. We have not shown the results since the plots are quite noisy
but we will be exploring this region further in subsequent work.

\subsection{Other statistics derived from the genus}

The genus at different threshold values is strongly
(anti)correlated. 
One may then think of deriving other
observables using the information inherent in the genus curves 
so that the non-Gaussian information may be maximized, and which may
distinguish between different kinds of non-Gaussianity. The
simulated non-Gaussian maps may then be used to test the sensitivity
of these observables. We mention here four such quantities, namely,
${\rm R_{cold}}$, ${\rm R_{hot}}$, ${\rm R_{spots}}$ and ${\rm S_0}$,
which are defined below. Then we explore
how they deviate from the Gaussian expectations. 

\begin{enumerate} 
\item ${\rm R_{cold}}$: Let ${\rm N_{cold}}$ denote the total number
  of cold spots, defined as  
${\rm N_{\rm cold}} \equiv \int_{-\nu_2}^{-\nu_1} d\nu\,  G(\nu), $
where  $\nu_1, \nu_2$ are suitably chosen positive threshold values
with $\nu_2>\nu_1$. Let $G^{\rm fit}(\nu)$ denote the Gaussian curve
obtained  by fitting the non-Gaussian genus points at different
threshold  values to a Gaussian shape. Let 
$ {\rm N^G_{\rm cold}} \equiv \int_{-\nu_2}^{-\nu_1} d\nu\,  G^{\rm
  fit}(\nu).  $
Then, we define 
\begin{equation}
{\rm R_{\rm cold}} \equiv \frac{{\rm N_{cold}}} {\rm N^G_{cold}}. 
\end{equation}
 For Gaussian maps, its value must be one. By inspecting
Figs.~(\ref{fig:pnge}) and choosing $\nu_1=1$ and $\nu_2=2.5$, 
we can predict that 
for $\gnl > 0$, ${\rm N_{cold}}$ must be less than one, whereas, 
for $\gnl < 0$ it must be greater than one. 

\item ${\rm R_{hot}}$: is defined to be the ratio ${\rm N_{hot}}/{\rm
  N^G_{hot}}$, where ${\rm N_{hot}}$ and ${\rm N^G_{hot}}$ are defined similar
  to ${\rm N_{cold}}$ and ${\rm N^G_{cold}}$, but for integration in the 
  positive threshold range.  Again the Gaussian expectation is one, 
and for same $\nu_1$ and $\nu_2$ as above ${\rm R_{  hot}}<1$ for
  $\gnl > 0$ and  ${\rm R_{ hot}}> 1$ for  $\gnl < 0$. If $\Delta
  G(\nu)$ is antisymmetric, as it appears to be in
  Fig.~(\ref{fig:pnge}), then we must have ${\rm R_{  hot}} = {\rm
  R_{cold}}$. 

\item  ${\rm R_{spots}}$: is defined to be the ratio ${\rm N_{spots}}
    /{\rm N^G_{  spots}}$, where ${\rm N_{spots}} = {\rm N_{cold}} +
    {\rm N_{hot}}$ and ${\rm N^G_{spots}} = {\rm N^G_{cold}} +
    {\rm N^G_{hot}}$. 
The Gaussian prediction of this statistic is one. 
For $\nu_1=1$ and $\nu_2=2.5$ it is less than one for $\gnl > 0$ and 
greater than one  for $\gnl < 0$.

\item ${\rm S}_0$ : is defined as the 
ratio of the slope of the non-Gaussian genus curve and that of the
fitted Gaussian at  $\nu=0$. It is greater than one for $\gnl > 0$ and
less than one for $\gnl < 0$. 
\end{enumerate}

\begin{figure}
\begin{center}
\includegraphics[height=0.3\textheight,width=0.45\textwidth]{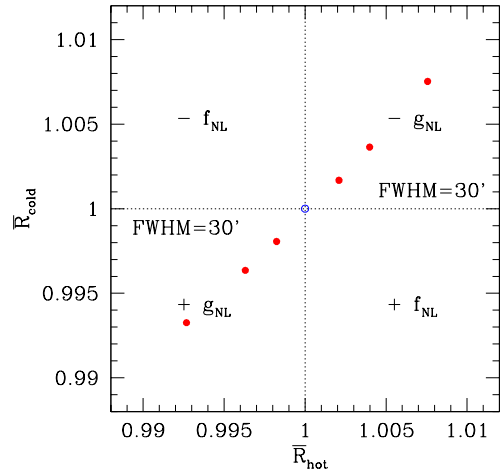}
\caption{${\rm {\overline R}_{hot}}$ {\em vs.} ${\rm {\overline R}_{cold}}$ 
for different values of $\gnl$ for ${\rm FWHM} = 30'$. The open circle
at the center  is
the Gaussian mean while filled ones are $\gnl=\pm 5\times 10^5, \pm 1\times
10^6$ and  $\pm 2\times 10^6$, at
increasing distances from (1,1). The region where the mean values for 
positive and negative values of $\fnl$  must lie, provided ${\rm
  FWHM}\lesssim 94'$,  are indicated. For ${\rm FWHM}\gtrsim 94'$, they
will interchange quadrants.}
\label{fig:hcmean}    
\end{center}
\end{figure} 

In the case of non-Gaussianity arising from $\fnl$, since the shape of
$\Delta G$ is strongly dependent on the smoothing angle, 
these observables will depend on the smoothing angle. 
By judicious choice of $\nu_1$ and
$\nu_2$  we can make them maximize the differences between $\fnl$ 
and $\gnl$ type non-Gaussianities. 
By inspecting Fig.~(2) of Ref.~\cite{Hikage:2006fe} and choosing
$\nu_1=1$ and $\nu_2=2.5$, one can deduce that
we must have 
${\rm R_{cold}} < 1$, ${\rm R_{hot}} >1$ for positive $\fnl$ and
${\rm R_{cold}} > 1$, ${\rm R_{hot}} <1$ for negative $\fnl$, 
when ${\rm FWHM \lesssim 94'}$ . The
situation is reversed if ${\rm FWHM \gtrsim 94'}$. 
${\rm S}_0$ is always greater than one for positive $\fnl$ and  always
less than one for negative case.
For ${\rm FWHM \lesssim 94'}$, ${\rm R_{ spots}} >1$ if $\fnl>0$ and  
${\rm R_{spots}} < 1$ if $\fnl<0$ and they interchange for  
${\rm FWHM \gtrsim 94'}$  . 

We next analyze the properties of these four observables inferred
from the non-Gaussian maps. An overhead bar denotes the mean value of
each of the above four statistics obtained by averaging over  200  
realizations.  Fig.~(\ref{fig:hcmean}) shows the parameter space  
$({\rm {\overline R}_{hot}},\, {\rm {\overline R}_{cold}})$ for 
varying $\gnl$ values, for FWHM$=30'$.  We have used
$\nu_1=1$ and $\nu_2=2.5$ to calculate them.  
The open blue triangle at $(1,1)$ indicates the Gaussian mean, 
while the red filled ones
denote $\gnl=\pm 5\times 10^5, \pm 1\times 10^6$ and   $\pm 2\times 10^6$, at
increasing distances from (1,1). In practice, the Gaussian mean has a
small shift from (1,1) due to statistical fluctuation. We have corrected
it by simply shifting it. We have then shifted the non-gaussian means
by the same amount. This shift does not affect the relative distance
between Gaussian and non-Gaussian means. 
Note that with minimal statistical fluctuation and if $\Delta G$ is
exactly anti-symmetric then the slope of the line $({\rm {\overline
    R}_{hot}},\, {\rm {\overline R}_{cold}})$  will be exactly
$45^{\circ}$. 

\begin{figure}
\begin{center}
\includegraphics[height=0.3\textheight,width=0.45\textwidth]
{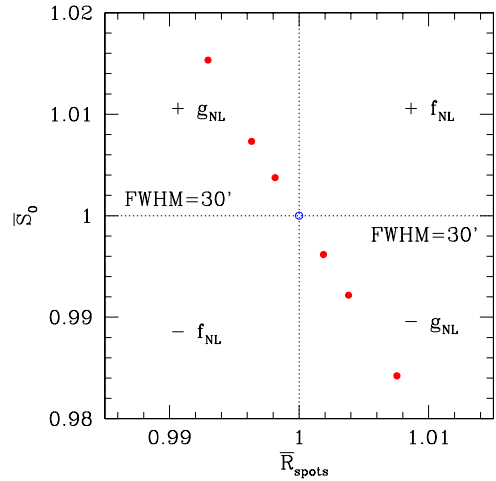}
\caption{${\rm {\overline R}_{spots}}$ {\em vs.} ${\rm {\overline
  S}}_0$  for different
  values of $\gnl$ for ${\rm FWHM} = 30'$. The open  circle at the 
center is
the Gaussian mean while the filled ones are $\gnl=\pm 5\times
  10^5,\ \pm 1\times
10^6$ and  $\pm 2\times 10^6$, at increasing distances from (1,1). 
The region where the mean values for 
positive and negative values of $\fnl$  must lie, provided ${\rm
  FWHM}\lesssim 94'$,  are indicated. For ${\rm FWHM}\gtrsim 94'$, they
will interchange quadrants.}
\label{fig:meanss}    
\end{center}
\end{figure} 

The parameter space 
$({\rm {\overline R}_{hot}},\, {\rm {\overline R}_{cold}})$ gets 
divided into four regions, with negative $\gnl$ occupying the first
quadrant and positive $\gnl$ the third, while $\mp \fnl$ occupies the
second and fourth quadrants, respectively. 
Larger smoothing angle increases the dispersion of the
distribution of individual points $({\rm  R_{hot}},\, {\rm 
R_{cold}})$ about the mean because the number of structures
decreases. But at the same time the size of deviations increases as
smoothing scale increases, making the low resolution maps also useful
in discriminating non-Gaussian maps from Gaussian ones.

Next we discuss the remaining two observables, ${\rm {\overline 
R}_{spots}}$ and ${\rm {\overline S}_{0}}$.  
Fig.~(\ref{fig:meanss}) shows the parameter space  
spanned by them for varying $\gnl$. 
$\nu_1$ and $\nu_2$ are the same as above.  
Again the blue triangle indicates the Gaussian mean, while the red ones
denote $\gnl=\pm 5\times 10^5, \pm 1\times 10^6$ and  $\pm 2\times 10^6$, at
increasing distances from (1,1). The smoothing scale is ${\rm
  FWHM}=30'$, and the effect of smoothing scale is similar to above. 
The parameter space gets
divided into four regions, with  $\pm\gnl$ occupying the second and
fourth quadrants, respectively, while $\pm \fnl$ occupies the
first and third quadrants, respectively.  Again for ${\rm FWHM}\gtrsim
94'$,  the positions of $+\fnl$ and $-\fnl$ will interchange.

\begin{figure}
\begin{center}
\includegraphics[height=0.3\textheight,width=.45\textwidth]{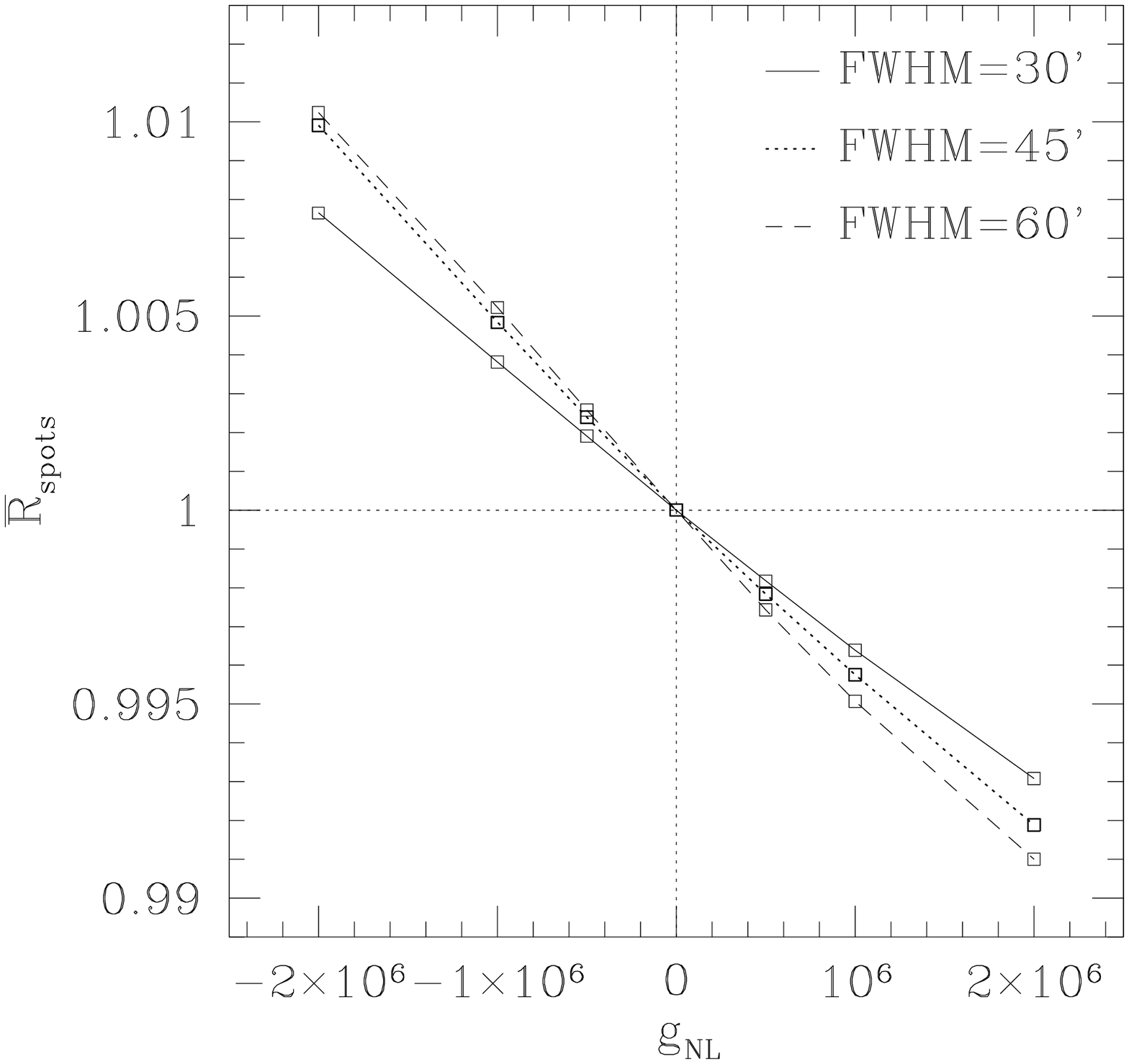}
\quad
\includegraphics[height=0.3\textheight,width=.45\textwidth]{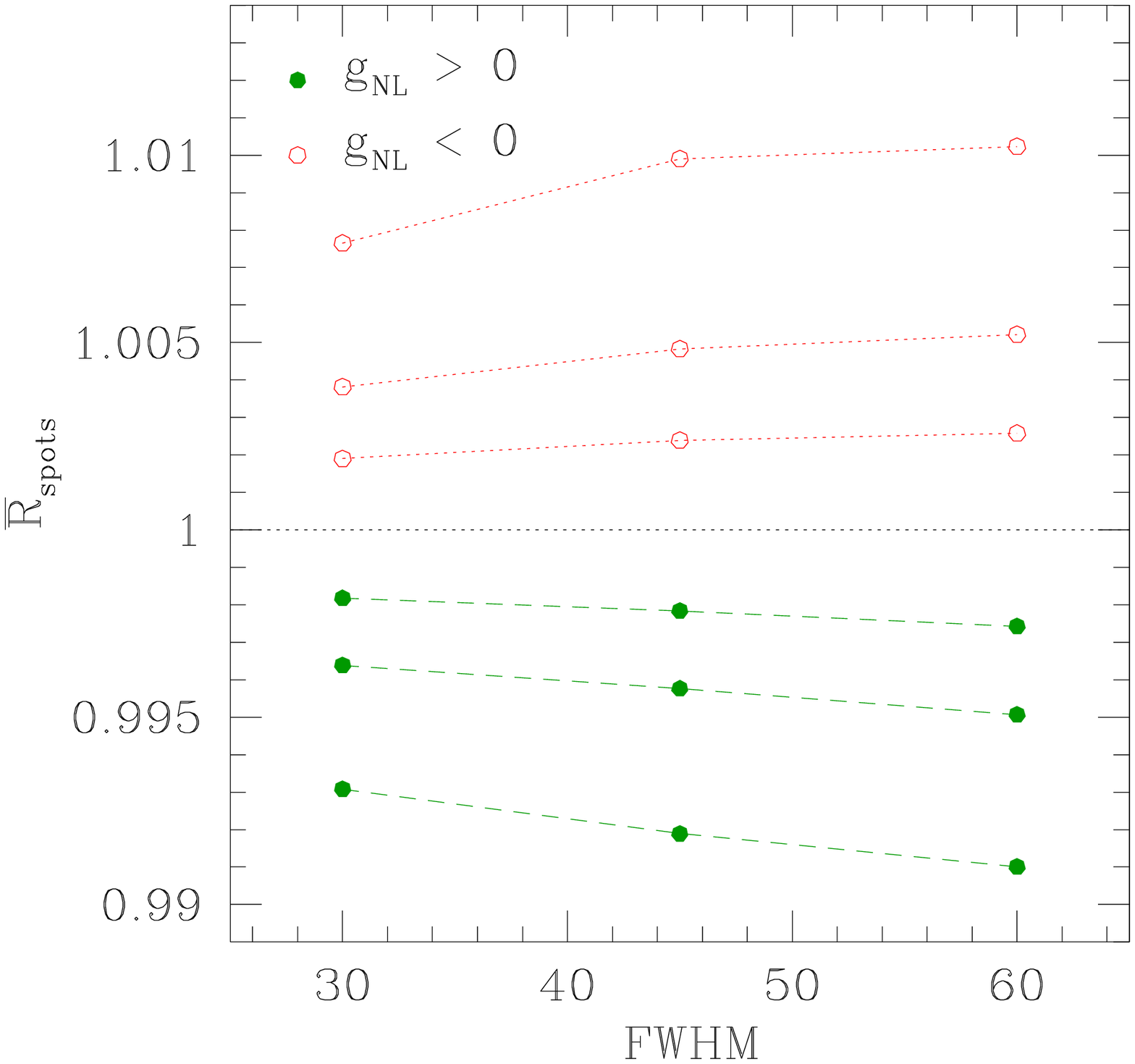}
\caption{Functional dependence of ${\rm R_{spots}}$ on $\gnl$
  and smoothing scale. In the right hand figure, the lines get farther
  away from ${\rm {\overline R}_{spots}}=1 $ line as $|\gnl|$
  increases. The values are $\gnl=\pm5\times 10^5, \pm10^6,\pm2\times 10^6 $. }
 \label{fig:spotsfunc}    
\end{center}
\end{figure} 

Since ${\rm  R_{spots}}$ carries information of both ${\rm  R_{hot}}$
and ${\rm  R_{cold}}$ we choose it for showing functional dependence
of $\gnl$ and FWHM. Panel 1 of Fig.~(\ref{fig:spotsfunc}) shows 
how ${\rm  R_{spots}}$ varies
with $\gnl$ when FWHM is fixed. As is clear from the figure we find
linear dependence, which agrees with the fact that we saw linear
dependence of $\overline{\Delta G}$ on $\gnl$ in
Fig.~(\ref{fig:dgbygfunc}). Panel 2 of Fig.~(\ref{fig:spotsfunc}) 
shows functional dependence on
FWHM for fixed $\gnl$. The lines get farther away from the Gaussian
expectation, ${\rm {\overline R}_{spots}} = 1$, as we increase
$|\gnl|$.   They exhibit mild increase with increase of the smoothing scale.

\section{Conclusion} 

We have simulated non-Gaussian CMB maps with the non-Gaussianity
coming from purely third order perturbations of the primordial
gravitational potential. We have used the map making algorithm proposed
by~\cite{liguori:2003} which computes $a_{\ell m}$'s as
an integral in real space. This method is particularly advantageous
for including the third order linearity, as compared to integrating in
Fourier space since there are two $k$ convolutions involved which make
the computational very heavy. 

We have investigated how the 1-point PDF gets modified from Gaussian
shape due to the effect of the non-linearity parameter $\gnl$. 
We found that positive $\gnl$ changes the 1-point PDF to leptokurtic
shape and negative $\gnl$ changes it to platykurtic shape.  
Its effect on the $C_{\ell}$'s is to increase power for large
$\ell$'s for both positive and negative $\gnl$ with the effect of
negative $\gnl$ being stronger.   

We have next used the simulated maps to compute the genus curve and
calculate their deviation from the Gaussian shape.    
The purpose is to understand their functional dependence on $\gnl$ and
how the non-Gaussian term modifies the topology of the CMB temperature
field. We found that positive $\gnl$ decreases both hot and cold spots
in the threshold range $1\lesssim |\nu| \lesssim 2.5$ and increases
the genus in the range  $0\lesssim |\nu| \lesssim 1$. The effect of
negative $\gnl$ is just the opposite. This results in antisymmetric
shapes of $\Delta G$, which look approximately like sine functions. 
We do not find significant variation of the
overall shape of $\Delta G$ as we vary smoothing scale, but the
amplitude of $\Delta G$ decreases as we increase the smoothing scale.   
We found that the sensitivity of the negative and positive $\gnl$'s
are roughly same. 
These results are very different from the genus arising from $\fnl$,  
which leads to symmetric form of $\Delta G$ and which has strong
dependence on the smoothing angle.  

We have also studied four other statistics derived from the genus, 
namely, the number of hot spots, the number of cold spots, 
the combined number of hot and cold spots and the 
slope of the genus curve at $\nu=1$. We found that these quantities 
carry distinct signatures of $\fnl$ and $\gnl$. 
 The parameter spaces get neatly divided into quadrants 
with each of the positive and negative $\gnl$ and  positive and
negative $\fnl$ occupying one. Hence they can be very useful for
distinguishing these two different types of non-Gaussianities. 

Since our goal was to get theoretical understanding of the nature of
non-Gaussianity arising from  $\gnl$ term by means of studying statistical
observables such as the genus and other quantities derived from it, 
we have not considered real observational contaminants such as point
sources, instrument noise etc., to our
simulations. One needs to take
them into account for actual comparison with experimental data and
putting constraints $\gnl$. Another interesting observable that we are
studying using the non-Gaussian simulations is the correlation of peaks
in the maps. These will be the subject of forthcoming publications.

\ack{The authors acknowlegde the support of the Korea Science and
  Engineering Foundation (KOSEF) through the Astrophysical Research
  Center for the Structure and Evolution of the Cosmos (ARCSEC). The
  computation in this paper was
  performed on the QUEST cluster at KIAS. P.C would like to thank
  Juhan Kim for his generous
  help while writing the codes, Qing-Guo Huang for many 
  useful discussions and Michel Liguori for useful communication. 
  We acknowledge use of the HEALPIX  and CMBFAST packages.}

\appendix
\section{Simplified expressions for $W_l$} 

Here we present expressions which simplify the numerical computation
of the filter function given in Eq.~(\ref{eqn:filter}).  
Instead of doing the integral over product of two $j_\ell$'s, which is
very time consuming due to the highly oscillatory behavior of
$j_\ell$, we can simplify
$W_\ell$ to express it in terms of Gamma functions. 
Using $j_\ell(kr) = \sqrt{\pi/2kr} J_{\ell+1/2}(kr)$ and 
$P_{\Phi}(k) = A_0 \ \frac{k^{n_s}} {k^4}$ we get,
\begin{equation}
W_{\ell}(r,r_1) = \sqrt{\frac{A_0}{r r_1}} \int_0^{\infty} dk\, k^{-1/2} 
J_{\ell+1/2}(kr)  J_{\ell+1/2}(kr_1)
\end{equation}

\noindent CASE 1 : $r=r_1$. Using the following formula~\cite{grad},
\begin{eqnarray}
 \int_0^{\infty} dk\ k^{-\lambda} J_{\nu}(\alpha k) J_{\mu}(\alpha k)
 &=& \frac{\alpha^{\lambda - 1}}{2^{\lambda}}  
     \frac{\Gamma(\lambda)} {\Gamma(\frac{-\nu + \mu+\lambda+1}{2})} 
\,\frac{ \Gamma(\frac{\nu + \mu-\lambda+1}{2})}  
{\Gamma(\frac{\nu + \mu+\lambda+1}{2}) } \nonumber\\
{}&{}& \times\, \frac{1}  {\Gamma(\frac{\nu - \mu+\lambda+1}{2})}, 
\end{eqnarray}
 which holds when $\nu + \mu +1 >  \lambda > 0$ and
 $\alpha > 0$, and putting $\lambda = 1-n_s/2$ we get,

\begin{equation}
W_\ell(r,r_1) = \frac{\sqrt{A}}{2^{1- \frac{n_s}{2}}} 
\frac{1}{r^{1+\frac{n_s}{2}}}  
\ \frac{\Gamma(1-\frac{n_s}{2})} {[\Gamma(1- \frac{n_s}{4} )]^2} \  
              \frac{\Gamma(\ell+ \frac{1}{2} +\frac{n_s}{4}      )} 
                 {\Gamma(\ell+ \frac{3}{2} - \frac{n_s}{4})}.
\end{equation}

\noindent CASE 2: $r \ne r_1$. For this case, when $\nu = \mu$, we can use the
following formula~\cite{grad},
\begin{eqnarray}
 \int_0^{\infty} dk\ k^{-\lambda} J_{\nu}(\alpha k) J_{\nu}(\beta k)
 {}&=& \frac{\alpha^{\nu} \beta^{\nu}}{2^{\lambda} 
(\alpha + \beta)^{2\nu -\lambda +1}}\,  
 \,\frac{\Gamma(\nu + \frac{1-\lambda}{2})}  
{\Gamma(\nu + 1) \,\Gamma(\frac{1+\lambda}{2}) } \nonumber \\
{} &{}& \times\, F\left( \nu + \frac{1-\lambda}{2}, \,\nu+1/2 ,
2\nu +1; \right.\nn\\
{}&{}& \left.\, \frac{4\alpha\beta}{(\alpha +\beta)^2} \right)
\end{eqnarray}
 which is holds provided $2\nu +1 >  \lambda > -1$ and
 $\alpha > 0,\ \beta>0$.  
$F(a,b;c;z) $ is the Gauss Hypergeometric function given by 
\begin{equation}
F(a,b;c;z) = \frac{\Gamma(c)}{\Gamma(b)\Gamma(c-b)} \int_0^1 dt \ 
             \frac{t^{b-1} (1-t)^{c-b-1}}{ (1-tz)^{a}}, 
\end{equation}
which is valid when $c > b >0$. Then, using  
\begin{equation}
\Gamma(2 \ell) = \frac{1}{\sqrt{2\pi}} 2^{2\ell-1/2} \,\Gamma( \ell)
\,\Gamma \left(\ell+ \frac{1}{2} \right)
\end{equation}
and defining  $z$ as 
\begin{equation}
z \equiv \frac{4rr_1} {(r+r_1)^2} =  \frac{4r_1/r} {(1+r_1/r)^2} , 
\end{equation}
we get 
\begin{eqnarray}
W_\ell(r,r_1) &=& \sqrt{\frac{A}{\pi}} \ 2^{n_s/2} 
               \frac{z^\ell}{(r+r_1)^{1+n_s/2}} \,
 \frac{ \Gamma(\ell+\frac{1}{2} + \frac{n_s}{4})} 
		   { \Gamma\left(1-\frac{n_s}{4}\right)\,\Gamma
               (\ell+1)} \nn\\
{}&{}& \times  \, \int_0^1 dt \frac{[ (1-t) t ]^\ell}
	      { (1-tz)^{\ell+\frac{1}{2}+\frac{n_s}{4} }}.
\end{eqnarray}
We can then scale the $r$ dependence as,
\begin{equation}
W_\ell(r,r_1) = \frac{1}{r^{1+n_s/2}}\, {\tilde W}_{\ell}(r_1/r),
\end{equation}
where
\begin{eqnarray}
{\tilde W}_{\ell}(r_1/r) &=& \sqrt{\frac{A}{\pi}} \ 2^{n_s/2} 
               \frac{z^\ell}{(1+r_1/r)^{1+n_s/2}}\  
\,  \frac{ \Gamma(\ell+\frac{1}{2} + \frac{n_s}{4})} 
   { \Gamma\left(1-\frac{n_s}{4}\right)\,\Gamma (\ell+1)}\nn\\
{}&{}&   \, \int_0^1 dt \frac{[(1-t) t]^\ell}
	       {(1-tz)^{\ell+ \frac{1}{2}+\frac{n_s}{4}}}. 
\label{eqn:wl}
\end{eqnarray}
Since $r\ne r_1$ we have $0\le z<1$. The integrand in
Eq.~(\ref{eqn:wl}) is a
smooth positive function with a local maxima in the interval $[0:1]$. It
becomes more and more localised as $\ell$ increases and the peak
position shifts towards one as $z$ approaches one. The integral can be
easily computed numerically.


\end{document}